\PassOptionsToPackage{pdfpagelabels=false}{hyperref} 
\documentclass[fleqn,usenatbib]{mnras}
 
\usepackage{newtxtext,newtxmath}
\usepackage[T1]{fontenc}

\usepackage{graphicx}	
\usepackage{amsmath}	

\usepackage{amsmath}
\usepackage{setspace}
\usepackage{widetext}

\usepackage{orcidlink}

\usepackage{eso-pic}

\AddToShipoutPictureBG*{%
  \AtPageUpperLeft{%
    \hspace{0.675\paperwidth}%
    \raisebox{-5.0\baselineskip}{%
      \makebox[0pt][l]{\textnormal{DES-2024-0830}}
}}}%

\AddToShipoutPictureBG*{%
  \AtPageUpperLeft{%
    \hspace{0.675\paperwidth}%
    \raisebox{-6.0\baselineskip}{%
      \makebox[0pt][l]{\textnormal{FERMILAB-PUB-24-0553-PPD}}
}}}%

\title[BCG and ICL Hierarchical Growth]{The Hierarchical Growth of Bright Central Galaxies and Intracluster Light as Traced by the Magnitude Gap}

\author[DES Collaboration]{
\parbox{\textwidth}{
\Large
Jesse~B.~Golden-Marx$^{1}$\orcidlink{0000-0002-6394-045X}\footnotemark,
Y.~Zhang$^{2}$\orcidlink{0000-0001-5969-4631},
R.~L.~C.~Ogando$^{3}$\orcidlink{0000-0003-2120-1154},
B.~Yanny$^{4}$\orcidlink{0000-0002-9541-2678},
M.~E.~S.~Pereira$^{5}$,
M.~Hilton$^{6,7}$,
M.~Aguena$^{8}$,
S.~Allam$^{4}$\orcidlink{0000-0002-7069-7857},
F.~Andrade-Oliveira$^{9}$,
D.~Bacon$^{10}$,
D.~Brooks$^{11}$\orcidlink{0000-0002-8458-5047},
A.~Carnero~Rosell$^{12,8}$\orcidlink{0000-0003-3044-5150},
J.~Carretero$^{13}$\orcidlink{0000-0002-3130-0204},
T.-Y.~Cheng$^{14}$\orcidlink{0000-0001-8670-4495},
L.~N.~da Costa$^{8}$,
J.~De~Vicente$^{15}$\orcidlink{0000-0001-8318-6813},
S.~Desai$^{16}$\orcidlink{0000-0002-0466-3288},
P.~Doel$^{11}$,
S.~Everett$^{17}$,
I.~Ferrero$^{18}$,
J.~Frieman$^{4,19}$\orcidlink{0000-0003-4079-3263},
J.~Garc\'ia-Bellido$^{20}$\orcidlink{0000-0002-9370-8360},
M.~Gatti$^{21}$,
G.~Giannini$^{13,19}$\orcidlink{0000-0002-3730-1750},
D.~Gruen$^{22}$\orcidlink{0000-0003-3270-7644},
R.~A.~Gruendl$^{23,24}$,
G.~Gutierrez$^{4}$\orcidlink{0000-0003-0825-0517},
S.~R.~Hinton$^{25}$,
D.~L.~Hollowood$^{26}$,
K.~Honscheid$^{27,28}$\orcidlink{0000-0002-6550-2023},
D.~J.~James$^{29}$\orcidlink{0000-0001-5160-4486},
K.~Kuehn$^{30,31}$\orcidlink{0000-0003-0120-0808},
S.~Lee$^{17}$,
J. Mena-Fern{\'a}ndez$^{32}$\orcidlink{0000-0001-9497-7266},
F.~Menanteau$^{23,24}$\orcidlink{0000-0002-1372-2534},
R.~Miquel$^{33,13}$\orcidlink{0000-0002-6610-4836},
J. Mohr$^{22}$\orcidlink{0000-0002-6875-2087}
A.~Palmese$^{34}$\orcidlink{0000-0002-6011-0530},
A.~Pieres$^{8,3}$\orcidlink{0000-0001-9186-6042},
A.~A.~Plazas~Malag\'on$^{35,36}$\orcidlink{0000-0002-2598-0514},
S.~Samuroff$^{37}$,
E.~Sanchez$^{15}$\orcidlink{0000-0002-9646-8198},
M.~Schubnell$^{9}$\orcidlink{0000-0001-9504-2059},
I.~Sevilla-Noarbe$^{15}$\orcidlink{0000-0002-1831-1953},
M.~Smith$^{38}$\orcidlink{0000-0002-3321-1432},
E.~Suchyta$^{39}$\orcidlink{0000-0002-7047-9358},
G.~Tarle$^{9}$\orcidlink{0000-0003-1704-0781},
V.~Vikram$^{40}$,
A.~R.~Walker$^{41}$\orcidlink{0000-0002-7123-8943},
N.~Weaverdyck$^{42,43}$,
and P.~Wiseman$^{44}$
\begin{center} (DES Collaboration) \end{center}
}
}

\date{Accepted for Publication in MNRAS February 2025}
\pubyear{2025}
\begin{document}
\label{firstpage}
\pagerange{\pageref{firstpage}--\pageref{lastpage}}
\maketitle

\begin{abstract}
Using a sample of 2800 galaxy clusters identified in the Dark Energy Survey across the redshift range $0.20 < z < 0.60$, we characterize the hierarchical assembly of Bright Central Galaxies (BCGs) and the surrounding intracluster light (ICL). To quantify hierarchical formation we use the stellar mass - halo mass (SMHM) relation, comparing the halo mass, estimated via the mass-richness relation, to the stellar mass within the BCG+ICL system. Moreover, we incorporate the magnitude gap (M14), the difference in brightness between the BCG (measured within 30\,kpc) and 4th brightest cluster member galaxy within 0.5 $R_{200,c}$, as a third parameter in this linear relation. The inclusion of M14, which traces BCG hierarchical growth, increases the slope and decreases the intrinsic scatter, highlighting that it is a latent variable within the BCG+ICL SMHM relation. Moreover, the correlation with M14 decreases at large radii. However, the stellar light within the BCG+ICL transition region (30\,kpc - 80\,kpc) most strongly correlates with halo mass and has a statistically significant correlation with M14. Since the transition region and M14 are independent measurements, the transition region may grow due to the BCG's hierarchical formation. Additionally, as M14 and ICL result from hierarchical growth, we use a stacked sample and find that clusters with large M14 values are characterized by larger ICL and BCG+ICL fractions, which illustrates that the merger processes that build the BCG stellar mass also grow the ICL. Furthermore, this may suggest that M14 combined with the ICL fraction can identify dynamically relaxed clusters. 

\end{abstract}

\footnotetext{E-mail: jesse.golden-marx@nottingham.ac.uk}
\section{Introduction}
\label{sec:intro}

Bright Central Galaxies (BCGs) are massive, radially extended, elliptical galaxies located near the centre of their host cluster's dark matter halo. BCGs grow hierarchically, by merging with smaller galaxies over time \citep[e.g.,][]{del07, bez09, naa09, ose10, van2010, wel06}. The combination of hierarchical formation and central location results in a correlation between the properties of the BCG and the cluster's dark matter halo \citep{jon84, rhe91, lin04, lauer14}. This galaxy-halo connection provides key information in characterizing the formation history of these massive galaxies.

Observations, simulations, and semi-analytic models \citep{cro06, del07, guo10, ton12, shankar15} suggest that BCGs grow as a result of a two-phase formation \citep{ose10, van2010}. At high redshift ($z > 2$) a dense core ($r < \approx$10kpc), which contains $\approx$25\% of the stellar mass is formed via in-situ star formation. At $z < 2$, the outer envelope grows hierarchically as a result of major and minor mergers. This outer envelope extends to the faint and diffuse halo of intracluster light \citep[ICL;][]{zwi33,zwi51} that is observed around BCGs. Indeed, the presence of ICL can be used to differentiate BCGs from similarly massive non-central galaxies \citep[e.g.,][]{hoe80, lyn00}. Therefore, it is unsurprising that prior works have suggested that at least part of the ICL forms due to the BCG's hierarchical assembly process \citep[e.g.,][]{mur07,gol23}.

Due to the faint and diffuse nature of the ICL, deep, high resolution photometry with a large field-of-view is needed to accurately characterize the ICL. Using such observations, individual clusters have measurements of the ICL that extend out to hundreds of kiloparsecs from the BCG \citep[e.g.,][]{klu20, gon21, mon21, mon22, gol23, klu24}. However, these observations can be enhanced by stacking the ICL signal surrounding many BCGs. Doing so, prior analyses have found that the ICL extends as far out as Mpc scales \citep{zib05,zha18,che22,zha24}. Although the ICL can be measured out to large radii, it is observationally challenging to distinguish the light associated with the BCG and that of the ICL \citep[see discussions in e.g.,][]{gon07, zha18, gon21, klu21, mon21, bro24}. Therefore, we do not disentangle the BCG and ICL; instead, we focus on the BCG+ICL system and use fixed radial apertures to define our measurements, as introduced in \citet{pil18}. In this analysis, we define similar radial regimes as \citet{zha24}; the BCG refers to the inner 30\,kpc, the BCG+ICL transition regime is 30\,kpc - 80\,kpc, and the ICL is the light beyond 80\,kpc. 

The stellar mass-halo mass (SMHM) relation is a commonly used observational formalism to quantify the galaxy-halo connection for galaxy clusters ($\rm log_{10}(\it{M}_{\rm halo}$ /$(M_{\odot}/h)) \ge 14.0)$ due to its low intrinsic scatter, $ \sigma_{\rm int}$, in stellar mass at fixed halo mass, which observations and simulations measure as $\approx$ 0.15 dex \citep[e.g.,][]{zu16,pil18, kra14,gol18}. In logarithmic scale, this is a linear relation, which directly compares the stellar mass of the BCG to the total mass of the cluster with the form: log$_{10}$(M$_{\rm *}$) $\propto$ slope $\times$ log$_{10}$(M$_{\rm halo}$). Moreover, measurements of the SMHM relation's parameters, such as the slope and $\sigma_{\rm int}$ provide insight into BCG growth and evolution \citep{gu2016, gol19,gol23}. 

As a result of the two-phase growth, information about the BCG's recent stellar mass growth is contained within the BCG's outer envelope and ICL \citep{ose10,van2010}. It follows that measurements of the SMHM relation in both observations and semi-analytic models have found that the slope increases in progressively larger radii \citep{mos18,gol19}. Moreover, \citet{gol23} extended this analysis to the ICL and found that the SMHM relation's slope increases to an asymptote at $\sim$100\,kpc. Moreover, recent work found that the underlying halo mass is more strongly correlated with the light from the outer envelopes (50\,kpc-100\,kpc \citep{hua21, kwi24} or 30\,kpc - 50\,kpc \citep{gol23}) than the light from the BCG's core. Thus, the ICL is imprinted with information about recent mergers, which strengthens the correlation with the underlying dark matter halo mass \citep[e.g.,][]{mon19}.  

Accounting for latent variables, which ideally increase the slope and decrease $\sigma_{\rm int}$ also strengthens statistical correlations. One observational measurement inherently tied to BCG hierarchical growth is the magnitude gap, the difference in $r-$band magnitude between the BCG and 4th brightest cluster member within half the radius enclosing 200 times the critical density of the Universe ($R_{\rm 200,c}$) \citep{dar10}, referred to as M14 throughout this analysis. N-body simulations found that BCG stellar mass linearly increases with the number of progenitor galaxies \citep{solanes16}. Thus, as BCGs grow hierarchically, they increase in stellar mass and brightness, while the satellite galaxies remain fixed or merge with the BCG. Therefore, BCG growth increases M14 yielding the correlation between M14 and BCG stellar mass \citep{har12,gol18,gol19,gol22}. Thus, M14 traces the BCG's hierarchical assembly \citep{gol18} and acts as an indicator of the BCG's dominance over other cluster member galaxies, similar to the Bautz-Morgan classification \citep{Bau70}. We also note that \citet{oli14} identify a similar offset in stellar mass at fixed halo mass between central and non-central BCGs, which they find have low magnitude gaps, defined using the 2nd brightest galaxy (M12), in galaxy groups and clusters from the GAMA survey. Additionally, \citet{vit18} found that clusters with larger M12 values, are more concentrated and likely formed earlier, suggesting that the magnitude gap traces cluster formation, which was also been found in N-body simulations of fossil group galaxies \citep{don05, von08}.

Based on \citet{gol18}, M14 is a statistical latent parameter within the cluster SMHM relation, as it increases the slope and decreases $\sigma_{\rm int}$ by as much as 50\% \citep{gol18,gol19,gol22}. Moreover, reducing $\sigma_{\rm int}$, tightens the constraints on the slope, which allowed for the detection of redshift evolution of the SMHM relation's parameters \citep{gol19,gol22}. We also note that the M14 correlation holds if M12 is used \citep{gol18}. We use M14 because it yields a stronger correlation with early formation \citep{dar10}. Moreover, the variance in the selection of the 4th brightest galaxy (and thus its magnitude) is less susceptible to large uncertainty due to foreground or background interloper galaxies, cluster mergers with multiple central galaxies, or recently infalling bright galaxies. 

The ICL and M14 result from BCG hierarchical assembly. However, it remains unclear how these parameters are related. The ICL is a direct measurement of the diffuse light that surrounds the BCG and can be used as a fossil tracer of the BCGs assembly history, while M14 is a measure of the BCGs hierarchical assembly within the cluster environment. Therefore, in this paper we present measurements of the correlation of the BCG+ICL stellar content with halo mass and M14 (an extension of the work presented by \citet{gol23}) to determine for the first time how these parameters correlate in the context of BCG hierarchical assembly and the SMHM relation. For this analysis, we use the Dark Energy Survey \citep[DES;][]{DES05} Y6 sample of clusters, currently the only statistically large (> 1000) sample of clusters that such measurements of the ICL can be taken within the redshift range $ 0.2 < z < 0.6$. 

The remainder of this paper is outlined as follows. In Section~\ref{sec:data}, we define our observational sample, the DES Y3 redMaPPer cluster sample and DES Y6 data, and the data reduction methods. In Section~\ref{sec:measurements}, we present the measurements of the stellar mass content of the BCG+ICL system, halo mass, and M14. In Section~\ref{sec:model}, we describe the hierarchical Bayesian model used to measure the parameters of the SMHM relation. In Section~\ref{sec:ObsSMHM}, we present our analysis of the SMHM relation for the BCG+ICL system. In Section~\ref{sec:M14_ICL}, we discuss the correlation between M14 and ICL, including the ICL fraction. Finally, we conclude in Section~\ref{sec:conclusion}. 

In this analysis, we assume a flat $\Lambda$CDM universe, with $\Omega_{M}$=0.30, $\Omega_{\Lambda}$=0.70, H$_{0}$=100~$h$~km/s/Mpc with $h$=0.7.

\section{data}
\label{sec:data}

\subsection{DES Y6 data}
\label{subsec:DESy6}
The Dark Energy Survey (DES) is a wide-area multi-band photometric sky survey covering $\approx$ 5,000 square degrees across five photometric wavebands ($g, r, i, z, Y$) taken using the Victor M. Blanco 4-m telescope beginning with Science Verification in 2012 and ending in 2019 with the Year 6 observations. The DES Year 6 data, which includes all prior DES data releases, features several layers of single-epoch images combined to create deep photometry that allow us to capture low surface brightness features. DECam, the imager used for DES, features a large 2.2 degree field-of-view, a 570 megapixel camera constructed using a CCD mosaic, and a low-noise electronic readout system \citep{flaugher2015}. These properties make DECAM ideal for studying faint and extended diffuse sources, including the ICL, in the optical from the ground in a single field-of-view. 

The data used throughout this analysis, both the source catalogs and the images, come from the DES Y6 co-addition (this includes publicly released data from DES Data Release 2 \citep[][]{DESDR2}). The DESY6 images are deeper than the Y3 versions and provide a more uniform co-add with an improved background subtraction. Following the prescription described in \citet{tan21}, the median surface brightness limits at 3$\sigma$ for the r- and i-band (the bands used in this analysis) are $r$ = $28.32^{+0.06}_{-0.12}$ and $i$ = $27.87^{+0.06}_{-0.09}$ mag arcsec$^{-2}$, which is $\sim$0.5 mag arcsec$^{-2}$ fainter than the DESY3 limits \citep{tan21}. Since this work focuses on the diffuse ICL, these improvements, which include the improved masking of faint galaxies located spatially near/within the ICL, are key. Such faint galaxies can unintentionally be treated as part of the ICL, thus, these improvements enhance our ability to accurately measure the contribution from the ICL by strengthening our ability to delineate between the diffuse light and total light within the cluster, which includes all cluster member galaxies and the ICL. We note that the improvements in the faint galaxy identification result from our use of the DES Y6 co-addition catalogues, which create a deeper catalogue by combining the full six years of r-,i-, and z-band photometry.  

The improvements seen in the final images presented in \citet{DESDR2} take advantage of additional exposures and tilings that cover the entire DES footprint. For these images, a single-epoch background subtraction is performed, instead of a ``global'' Swarp \citep{2010ascl.soft10068B} background subtraction. Like in \citet{gol23} and \citet{zha24} the science images are an average of the $riz$ photometry. Moreover, the DESY6 photometry allows for a fainter detection threshold (5$\sigma)$ than what was used in previous data releases (10$\sigma$).

\subsection{redMaPPer Cluster Sample}
\label{subsec:redMaPPer}
In this analysis, we use the DES Y3 redMaPPer cluster catalog version 6.5.22+2, based on DES Year 3 Gold photometric data \citep{sev21}. This version of the redMaPPer cluster catalogue contains over 21,000 galaxy clusters with a redMaPPer defined richness, $\lambda$, greater than 20, which corresponds to a halo mass lower limit of $10^{14.1}$M$_\odot$ \citep{mcc19}. Moreover, we take advantage of redMaPPer's membership catalogue which provides the probability ($P_{\rm mem}$) and DES r-band model magnitude of each cluster member, to measure M14 and central galaxy probability ($P_{\rm Cen}$), for BCG identification. We note that redMaPPer selects a luminous cluster galaxy located nearest to the centre of the cluster's gravitational potential well and provides the information for the five most probable central candidates per cluster. We assume that the most likely central galaxy candidate is the BCG. Based on multi-wavelength measurements, this selection has been shown to be correct with an $\sim 80\%$ frequency \citep{sar15, zha19, ble20}.

\subsection{Measuring the ICL}
\label{subsec:ICL_measurement}
The BCG and ICL profiles used in this analysis are data products derived as part of \citet{zha24}, which uses a similar methodology as \citet{zha18}, \citet{san21}, \citet{gol22}, and \citet{gol23}. Below, we briefly summarize the methodology used in \citet{zha24}. 

\begin{itemize}
    \item We analyse only clusters in the redshift range 0.2 $< z <$ 0.6, which reduces the number of redMaPPer ($\lambda >$ 20) clusters from 21092 to 15654. This redshift range was selected due to the completeness range of DES-redMaPPer \citep{ryk16}.
    \item The BCGs were selected as the most probable central galaxy provided by redMaPPer. The impact of using all BCGs regardless of redMaPPer $P_{\rm Cen}$ is discussed in the Appendix.
    \item Using these BCGs, the DES database was queried to select images within a 0.15 deg $\times$ 0.15 deg region centred on the BCG \footnote{For the redshift range in this study 0.15 deg ranges from 1.7Mpc - 3.6Mpc}. These individual images were combined (using mean pixel values) to create the final coadd image centred on the BCG. The images used throughout this analysis have not had a local sky background subtraction (measured either with Source Extractor \citep{1996A&AS..117..393B} or Swarp \citep{2010ascl.soft10068B}) applied. Instead, these images underwent a global background subtraction process \citep[estimated across the whole Field-of-View, about 2.5 $\mathrm{deg}^2$ for single exposure images;][]{2017PASP..129k4502B}. This background subtraction is done because local background regions often include the ICL, which results in an underestimation of the light in the BCG+ICL outer profiles \citep[e.g.,][]{von07, gol18}.
    \item Following the process described in \citet{zha18}, we remove clusters with BCGs located within 2000 DES pixels (526\arcsec) from bright foreground stars and large nearby galaxies. This was done to remove saturated stars and yield a better sky estimate. These objects were identified using the bad region mask defined in \citet{drl18}. Removing these clusters reduces the number of available clusters in our sample to 7042.
    \item Using the BCG as the centre, all objects covering a 0.2 deg $\times$ 0.2 deg region surrounding the BCG in the DES database were identified. We acknowledge that miscentring exists in redMaPPer (see \citet{zha19}). However, this is not corrected for in the reduction process. As shown in the Appendix, when we look at high and low probability centrals, we find no evidence that it impacts the parameters of the SMHM relation as shown in Tables~\ref{tab:SMHM_Posteriors_PCenlow} and \ref{tab:SMHM_Posteriors_PCenhigh}. 
    \item Elliptical masks with a semi-major axis of 3.5 $R_{\rm Kron}$\citep{Kron80}, were placed around all galaxies above a selected masking limit) excluding the BCG, allowing for the measurement of a BCG+ICL profile. The masking limit is a $z$-band magnitude limit, given in Table~\ref{tab:maglim}, based on the cluster's photometric redshift, of 0.2$L*$, where $L*$ is the characteristic luminosity of a cluster red galaxy luminosity function measurement \citep{zha19b}. We note that coadd catalogues are generally complete above 23.7 magnitudes in the z-band, so this tighter limit is done out of an abundance of caution. 
\begin{table}
\centering
\caption{Satellite Galaxy Masking Limits}
\begin{tabular}{cc}
\hline
redshift & z-band magnitude limit\\
\hline
0.2 $< z <$0.3  & 20.7\\
0.3 $< z <$0.4 & 21.4\\
0.4 $< z <$0.5 & 21.9\\
0.5 $<z <$0.6 & 22.3\\
\hline
\end{tabular}
\label{tab:maglim}
\end{table}
    
    \item Using the BCG as the central point, the surface brightness profiles (mean values) of the unmasked regions for each galaxy cluster, representative of the BCG+ICL light profile, were measured in circular radial annuli. We note that using simulations, \citet{bro24} find no statistical difference between measurements of the ICL done using circular and elliptical annular profiles. These measurements are the surface brightness radial profile for each cluster, which contain light from the cluster's BCG and ICL as well as some background light. Masked regions are excluded from the measurement process. 
    \item The surface brightness measurements described in the previous step also contain residual background light and contributions from unmasked foreground and background objects. For each radial surface brightness measurement, we take their values at the radial range beyond 500\,kpc as the ``background'' value, and subtract this ``background'' value from the corresponding measurements. The robustness of this background measurement was discussed in the Appendix of \citet{gol23}.
    \item Lastly, we remind the reader that unlike \citet{zha18}, \citet{san21}, and \citet{zha24}, we do not stack the measurements of the ICL for a large sample of clusters. Instead, we only combine images of the same cluster to create a deeper photometric image. 
\end{itemize}

\section{The Observable Measurements for Constructing the SMHM Relation}
\label{sec:measurements}

\subsection{BCG and ICL Spatial Definitions}
As noted in Section~\ref{sec:intro}, it is observationally difficult to disentangle the light in the BCG from the ICL. Although we focus on the combined BCG+ICL light profile, for the purposes of the SMHM relation, we use the fixed radial apertures introduced in \citet{zha24} to separate the BCG and ICL into 3 regions, 0-30\,kpc (the BCG), 30-80\,kpc (the transition region), and beyond 80kpc (the ICL). This choice of radial ranges is in excellent agreement with the individual ICL studies of Abell 85 \citep{mon21} and the Perseus cluster \citep{klu24}.

\begin{figure}
    \centering
    \includegraphics[width=8cm]{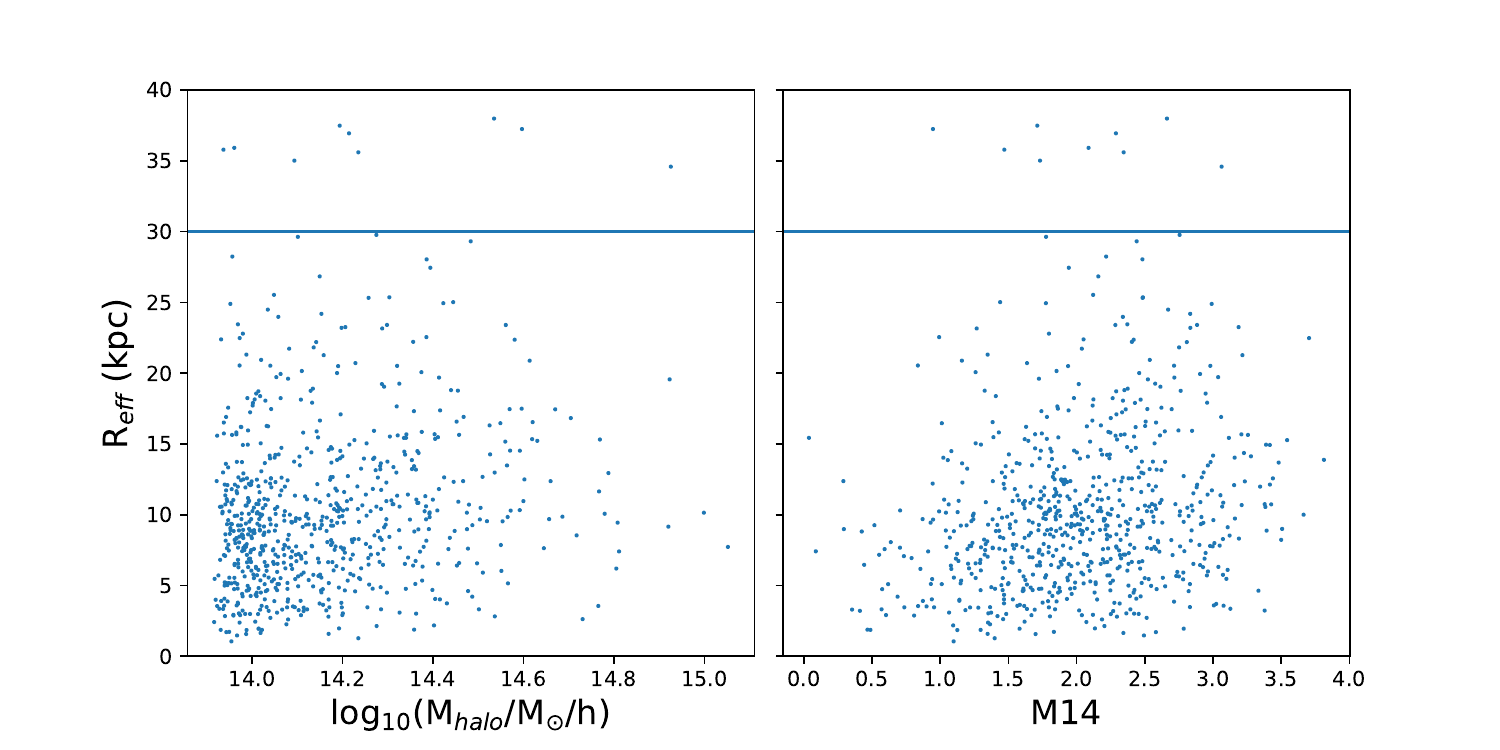}
    \caption{The effective radius, estimated from a S\'ersic fitting, plotted against the halo mass and M14.  We see that our choice of 30\,kpc to define the BCG includes two times the effective radius for the vast majority of BCGs. Moreover, we find no trends in the effective radius with either halo mass or effective radius.}
    \label{fig:sersic}
\end{figure}

To further verify our choice for the inner region of the BCG, we fit the inner portion of our light profile to a single S\'ersic profile, allowing us to measure the effective radius, R$_{\rm eff}$, of the BCG. In Figure~\ref{fig:sersic}, we compare the effective radius to both the halo mass estimated using redMaPPer richness (described in Section~\ref{subsec:Mass}) and to M14 (described in Section~\ref{subsec:DESmgap}). Though BCG size scales with halo mass, we see no trend in the effective radius with either halo mass or M14 due to the high mass range of our sample. Moreover, our selection of 30\,kpc encapsulates two times the effective radius for the vast majority of our sample, which suggests that our choice of apertures generally captures the light of the BCG and ICL, respectively.

\subsection{Stellar Mass}
The stellar mass is estimated in the same manner as in \citet{gol23}, so we summarize the method. For each cluster, using the BCG+ICL system's light profile (see Section~\ref{subsec:ICL_measurement}) we measure the apparent magnitude within a given radial aperture by integrating the light profile between the selected inner and outer radii in the DES i-band. In this analysis we use the following apertures: 0-30\,kpc, 0-80\,kpc, 0-150\,kpc, 0-225\,kpc, 0-300\,kpc, 30-80\,kpc, 80-150\,kpc, 150-225\,kpc, 225-300\,kpc. 

Using the selected aperture magnitudes, we use the EzGal \citep{man12} SED modelling software to estimate the stellar mass. For EzGal, we assume a passive spectral model, as DES photometry prevents us from statistically constraining additional star formation parameters such as burst times and formation epochs. For converting magnitude to stellar mass, we assume the same parameters as \citet{gol19}, \citet{gol22}, and \citet{gol23}; a \citet{bru03} stellar population synthesis model, a \citet{sal55} Initial Mass Function (IMF), a formation redshift of $z=4.9$, and a metallicity of 0.008 (66\%~Z$_{\odot}$). These parameters were selected to minimize the $\chi^2$ statistic between the measured and EzGal modeled photometry. Following \citet{gol22} and \citet{gol23}, we assume a subsolar metallicity for each BCG+ICL system since observations find that the metallicity of early type galaxies decreases radially outward \citep[e.g.,][]{lou12, mcd2015, oli15, edw24} and light from the ICL has been characterized as having a subsolar metallicity \citep{mon18}.  Additionally, our choice of a passive spectral model is supported by observations which show that both BCGs \citep[e.g.,][]{mcd16} and the ICL \citep[e.g.,][]{mcc23} lack recent star formation. We also note that our stellar mass estimate is independent of our choice of formation redshift and that changing the \citet{sal55} IMF to a \citet{cha03} IMF uniformly lowers our stellar mass estimates by $\approx$~0.25 dex. Thus, the SMHM relation's amplitude is the only SMHM parameter impacted. 

Since our Bayesian infrastructure relies on an estimate of measurement uncertainty, we assume the same uncertainty in stellar mass, 0.06 dex, as in \citet{gol22} and \citet{gol23}. This is likely a lower limit on the estimated uncertainty in stellar mass, in particular when we incorporate the ICL. If a larger uncertainty is required, this only changes $\sigma_{\rm int}$. The remaining parameters are consistent. While our MCMC analysis measures $\sigma_{\rm int}$, a detected change of $\sigma_{\rm int}$ with radius may reflect our underestimation of the uncertainty in the stellar mass at large radii.

\subsection{DES Cluster Richnesses and Halo Masses}
\label{subsec:Mass}
Unlike in \citet{gol23}, which used the DES-ACT overlap sample \citep{hil20} and SZ-estimated halo masses, for the DES clusters, we use redMaPPer richness, $\lambda$. This richness is converted to halo mass using the DES Year 1 calibrated mass-richness relation \citep{mcc19}, given by Equation~\ref{eq:DESm-lambda},
\begin{equation}
    \label{eq:DESm-lambda}
    M_{\rm halo}/(\it h^{-1} M_{\odot}) = \rm{10^{14.344}} (\lambda/40)^{1.356} (\frac{1+\it{z}_{\rm red}}{1+0.35})^{-0.30}
\end{equation}
where $M\rm_{\rm halo}$ refers to $M\rm_{200m}$, $z\rm_{red}$ is the redMaPPer photometric redshift and $\lambda$ is the redMaPPer richness. We note that the \citet{mcc19} mass-richness relations has an intrinsic scatter associated with the halo mass at fixed richness. While not shown in Equation~\ref{eq:DESm-lambda}, we account for this scatter in our Bayesian MCMC analysis, as discussed in Section~\ref{sec:model}. Although we are using the DES Y3 redMaPPer catalogue a preliminary analysis of the DES-redMaPPer Y3 mass-richness relationship is consistent with the Y1 analysis. 

We note that a bias has been found in the weak lensing calibration of the \citet{mcc19} mass-richness relation. The halo masses are biased $\approx$0.1 dex low, with the $\lambda <$30 clusters most strongly impacted \citep{DES2020, cos21}. For this analysis, if only the low-$\lambda$ clusters have halo masses that are biased low, this bias would result in a value of the slope that is also biased low and may also impact the offset parameter. Since this would be a uniform bias across each aperture measurement of the SMHM relation, the bias would not affect our ability to characterize how incorporating ICL and M14 change the SMHM relation and thus would not impact the conclusions of this analysis. 

Here, we treat the uncertainty in the mass-richness relation as a fixed value identical to what was found for the SDSS and DES redMaPPer clusters in \citet{gol19} and \citet{gol22} because these measurements come from the same algorithm. We remind the reader that our chosen value, 0.087 dex, was determined in \citet{gol19} from a joint analysis where the parameters for the SMHM relation were simultaneously determined for a sample with halo masses estimated by richness and the caustic phase-space technique. This scatter in halo mass at fixed richness corresponds to 0.20 in a natural log scale, which is in excellent agreement with \citet{rozo2015}, which measures a value between 0.17-0.21.

\subsection{Magnitude Gap (M14)}
\label{subsec:DESmgap}
M14 is defined uniformly throughout this paper, regardless of our choice of BCG+ICL aperture stellar mass. For the DES-redMaPPer clusters we identify cluster members as those with $P_{\rm mem} \ge 0.9$. This is less restrictive than \citet{gol19} and \citet{gol22} since we do not match to the C4 catalog \citep{mil05}. This $P_{\rm mem}$ choice optimizes the number of clusters in our sample and only reduces the sample from 7042 to 6903 clusters. When we did this analysis with a threshold of $P_{\rm mem} \ge 0.8$ our results were consistent. 

We define M14 following \citet{dar10} as the difference in the $r-$band apparent model magnitude of the 4th brightest cluster member (with $P_{\rm mem} \ge $0.9) within 0.5$R_{\rm 200,c}$ and the BCG's inner 30\,kpc $r-$band apparent magnitude. We estimate $R_{\rm 200,c}$ for redMaPPer clusters using the following equations from \citet{ryk14}:
\begin{equation}
    \label{eq:DES_R200}
    R_{200,c} \approx 1.5 R_{c}(\lambda)
\end{equation}
where $\lambda$ is the redMaPPer richness, and $R_{c}$ is the redMaPPer cutoff radius, given by Equation~\ref{eq:DES_Rc}:
\begin{equation}
    \label{eq:DES_Rc}
    R_{c}(\lambda) = 1.0 h^{-1} (\lambda/100)^{0.2} \rm{Mpc}.
\end{equation}

Since we use DES data, we estimate the M14 uncertainty in the same manner as in \citet{gol22} and assume a value of 0.31. While this value may be an upper limit due to our lower $P_{\rm mem}$ criterion (as opposed to $P_{\rm mem} >$0.984 used in \citet{gol22}) this is unlikely to change our results as our Bayesian analysis is robust to moderate changes in the M14 uncertainty. For example, if we increase the M14 uncertainty to 0.5, our values remain in agreement. 

\begin{figure}
    \centering
    \includegraphics[width=7.5cm]{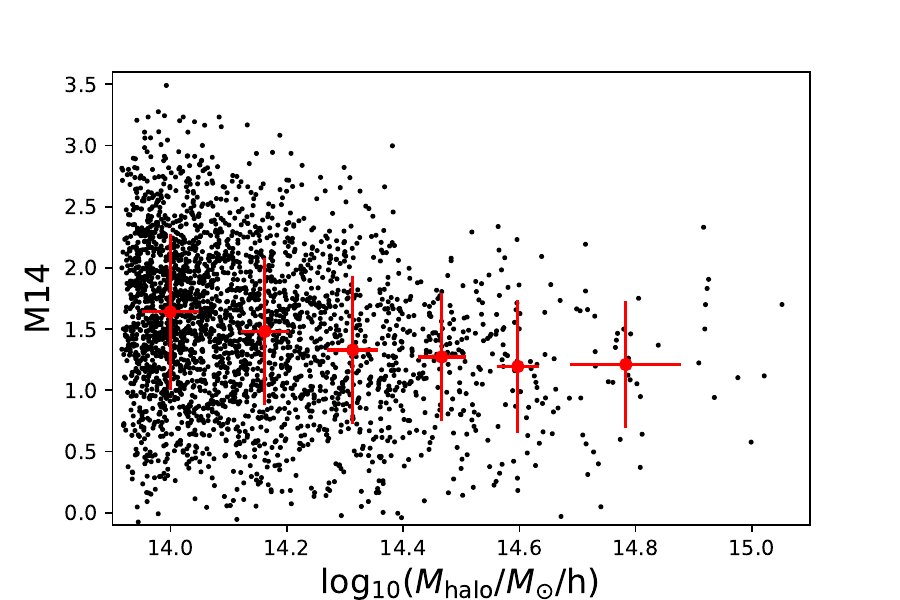}
    \caption{The M14 values plotted against the halo mass values (in black). Using 6 bins, we measure the median and standard deviation for both M14 and halo mass (in red) and find that the median value of M14 does not change.}
    \label{fig:M14_Mhalo}
\end{figure}

In Figure~\ref{fig:M14_Mhalo}, we plot the M14 values against the DES-redMaPPer halo masses (estimated using Equation~\ref{eq:DESm-lambda}). Using 6 halo mass bins, we measure the median values of M14 as a function of halo mass (in red). The median value is consistent (within less than 1$\sigma$) across the entire range in halo mass. We note that the upper limit on M14 does slightly decrease with halo mass; however, this is likely a result of small number statistics and hierarchical growth (i.e., the most massive clusters are later forming \citep[e.g.,][]{mat17} and still evolving.)

\section{The Statistical Modelling of the SMHM Relation}
\label{sec:model}

We measure the parameters of the SMHM relation using a similar hierarchical Bayesian MCMC approach to what is described in \citet{gol19} and \citet{gol23}. We quantify the SMHM relation using Equation~(\ref{eq:SMHM_M14}), when M14 is incorporated
\begin{equation}
\label{eq:SMHM_M14}
    \rm{log}_{10}(M_{*}/{M}_{\odot})=\textit{N}(\alpha + \beta \times \rm{log}_{10}(M_{\rm halo}/{M}_{\odot}) + \gamma \times (M14),\sigma_{\rm int}^2)
\end{equation} 
and Equation~\ref{eq:SMHM}, when it is not
\begin{equation}
\label{eq:SMHM}
    \rm{log}_{10}(M_{*}/{M}_{\odot})=\textit{N}(\alpha + \beta \times \rm{log}_{10}(M_{\rm halo}/{M}_{\odot}),\sigma_{\rm int}^2)
\end{equation} 
where \textit{N} refers to a normal distribution, $\alpha$ is the mathematical offset, $\beta$ is the slope of the SMHM relation, $\gamma$ is the M14 parameter introduced in \citet{gol18}, and $\sigma_{\rm int}$ is the intrinsic scatter in stellar mass at fixed $M_{\rm halo}$. We note that $\gamma$ measures the slope between the stellar mass and M14. Physically, $\gamma$ is related to hierarchical growth, such that a non-zero $\gamma$ means that the associated stellar mass is growing at least partially hierarchically. Also, the significance of this detection is given by how many standard deviations the value of $\gamma$ is from 0. This is a nested model, if $\gamma$=0, Equation~\ref{eq:SMHM_M14} reverts to Equation~\ref{eq:SMHM}. The only differences between the model used in this analysis and those in \citet{gol19} and \citet{gol23} are that we do not account for redshift evolution and we include $\gamma$, respectively.

A Bayesian formalism works by convolving prior information associated with a selected model with the likelihood of the observations given that model, to yield the posterior distribution for the non-nuisance parameters in our model. To determine the posterior distributions for each SMHM relation parameter, our MCMC model generates values for the observed aperture stellar masses, richness-estimated halo masses, and M14s at each step in our likelihood analysis, which are directly compared to the observed measurements.

We model the log$\rm_{10}$ BCG stellar masses ($y$), log$\rm_{10}$ halo masses ($x$), and M14 values ($z$) as normal distributions with mean values (locations) taken from our measurements. The standard deviations associated with each value is taken from measurement uncertainties, described in Section~\ref{sec:measurements} ($\sigma_{x0}$, $\sigma_{y_{0}}$, $\sigma_{z_{0}}$) as well as a stochastic component from a beta function (Beta(0.5,100)) \citep{gol18}, which allows for additional uncertainty on the observational errors. These errors are treated statistically in the Bayesian model as free nuisance parameters $\sigma_{x}$, $\sigma_{y}$, and $\sigma_{z}$. Additionally, as done in \citet{gol19}, \citet{gol22}, and \citet{gol23}, we reduce the covariance between $\alpha$ and $\beta$ by subtracting off the median values of the upper and lower limits in log$_{10}(M_{*}$/$M_{\odot})$ and log$_{10}(M_{\rm halo}$/$M_{\odot})$, 11.3 and 14.55, respectively. For consistency we subtract off the median stellar mass measured within 0-30\,kpc in each aperture bin. 

Due to the lack of detected evolution in the stellar mass of the ICL found in \citet{gol23} and \citet{zha24}, and the results from \citet{gol22}, which find that the majority of the evolution in the parameters of the SMHM relation (when including M14) occurs in the redshift range $z < 0.25$, much of which is below the redshift range we observe, we do not allow for the parameters of the SMHM relation to vary with redshift.

The parameters of the SMHM relation are given in Table~\ref{tab:bayes}, where each $i^{th}$ cluster is a component in the summed log likelihood and the terms marked $0i$ are representative of the observed data. We express the entire posterior as:

\begin{equation}
\begin{aligned}
p(\alpha,\beta,\gamma,\sigma_{\rm int}, x_{i},\sigma_{y_i},\sigma_{x_i},\sigma_{z,i}| x,y,z) \propto  \\
 \underbrace{\sum_{i} P(y_{0i}|\alpha,\beta,\sigma_{y_i},\sigma_{\rm int}, x_{i}, z_{i}) ~ P(x_{0i}|x_{i},\sigma_{x_i}) ~ P(z_{0i}|z_{i},\sigma_{z_i})}_{\text{likelihood}} \times \\  
 \underbrace{p(x_i) ~ p(\sigma_{x_i}) ~ p(\sigma_{y_i}) ~ p(\sigma_{x_i}) ~ p(\alpha) ~ p(\beta) ~ p(\gamma) ~ p(\sigma_{\rm int})}_{\text{priors}} 
\end{aligned}
\label{eq:DES-ACTposterior}
\end{equation}

Additionally, we reiterate that this is a nested Bayesian model, which allows for $\gamma$ to equal 0. Thus any statistically significant non-zero $\gamma$ is real and signifies that including M14 improves the fit to the measured data. If a correlation between M14 and stellar mass did not exist in our data or added additional noise, we would not detect a statistically significant $\gamma$ parameter. This discussion is expanded in Table~\ref{tab:SMHM_Posteriors_shuffled} in the Appendix.

\begin{table*}
\centering
\caption{Bayesian Analysis Parameters for the DES Sample}
\begin{tabular}{ccc}
\hline
Symbol & Description & Prior\\
\hline
$\alpha$ & The offset of the SMHM relation & $\mathcal{U}$(-20,20) \\
$\beta$ & The high-mass power law slope & Linear Regression Prior \\ 
$\gamma$ & The stretch parameter which describes the stellar mass - M14 stratification & Linear Regression Prior \\
$\sigma_{\rm int}$ & The uncertainty in the intrinsic stellar mass at fixed $M_{\rm 200m}$ & $\mathcal{U}(0.0,0.5)$\\ 
$y_{i}$ & The underlying distribution in stellar mass & Deterministic (inferred from Equation~\ref{eq:SMHM_M14}) \\ 
$x_{i}$ & The underlying $M_{\rm 200m}$ distribution & $\mathcal{N}$(14.11,$0.21^2$)\\
$z_{i} $ & The underlying M14 distribution & $\mathcal{N}$(1.74,$0.66^2$)\\ 
$\sigma_{y_{0i}}$ & The uncertainty between the observed stellar mass and intrinsic stellar mass distribution & 0.06 dex\\ 
$\sigma_{x_{0i}}$ & The uncertainty associated with log$_{10}$($M_{\rm 200m}$) &  0.087 dex  \\
$\sigma_{z_{0i}}$ & The uncertainty between the underlying and observed M14 distribution & 0.31 dex \\
\hline
\end{tabular}
\small
\\
\label{tab:bayes}
\end{table*}

\section{The Observed SMHM Relation}
\label{sec:ObsSMHM}

As discussed in Section~\ref{sec:intro}, many recent analyses have focused on the connection between the ICL and the cluster's dark matter halo. Recent observations \citep[e.g.,][]{mon19} suggest the light distribution of the ICL may trace the halo's dark matter distribution. Moreover, prior results have identified statistically significant correlations between the stellar mass contained within the ICL and the cluster's halo mass \citep[e.g.,][]{san21,hua21, gol23, kwi24}. As presented in \citet{gol23}, accounting for the ICL increases the slope, $\beta$, of the SMHM relation by 0.1-0.15. Thus, as a result of BCG+ICL hierarchical growth, the larger slope is measured because ICL, due to recent growth, is more correlated with the dark matter halo mass. Here, for the first time, we analyse the correlation between M14 and the ICL and investigate both quantitatively and qualitatively the impact of incorporating the ICL and M14 into the SMHM relation.

Our observational sample consists of all redMaPPer clusters ($\lambda$ > 20) in the redshift range $0.2 < z < 0.6$ with at least 4 high probability member galaxies ($P_{\rm mem}$ > 0.90). We remove a large fraction of clusters that as a result of masking of either bright stars or nearby galaxies, we are unable to measure the BCG+ICL profile, reducing our sample size to 2929. Lastly, we remove those clusters that we were unable to measure the light out to 300\,kpc leaving our final analysis sample as 2788 cluster. These criteria are summarized in Table~\ref{tab:sample}.

\begin{table}
\centering
\caption{Sample Selection}
\begin{tabular}{cc}
\hline
Criteria & Number of Clusters\\
\hline
$\lambda$ > 20 redMaPPer sample & 21092\\
0.2 $< z <$ 0.6 & 15654\\
No nearby bright foreground objects & 7042\\
4 $P_{\rm mem} > $ 0.9 members & 6903\\
ICL unimpacated by masking of nearby galaxies & 2929\\
Can measure ICL profile out to 300\,kpc & 2788\\
\hline
\end{tabular}
\\
\label{tab:sample}
\end{table}

\subsection{The BCG+ICL+M14 SMHM relation}
\label{subsec:BCG_ICL_SMHM}
\begin{figure*}
    \centering
    \includegraphics[width=20cm, trim={3cm 3.5cm 1cm 3.5cm},clip]{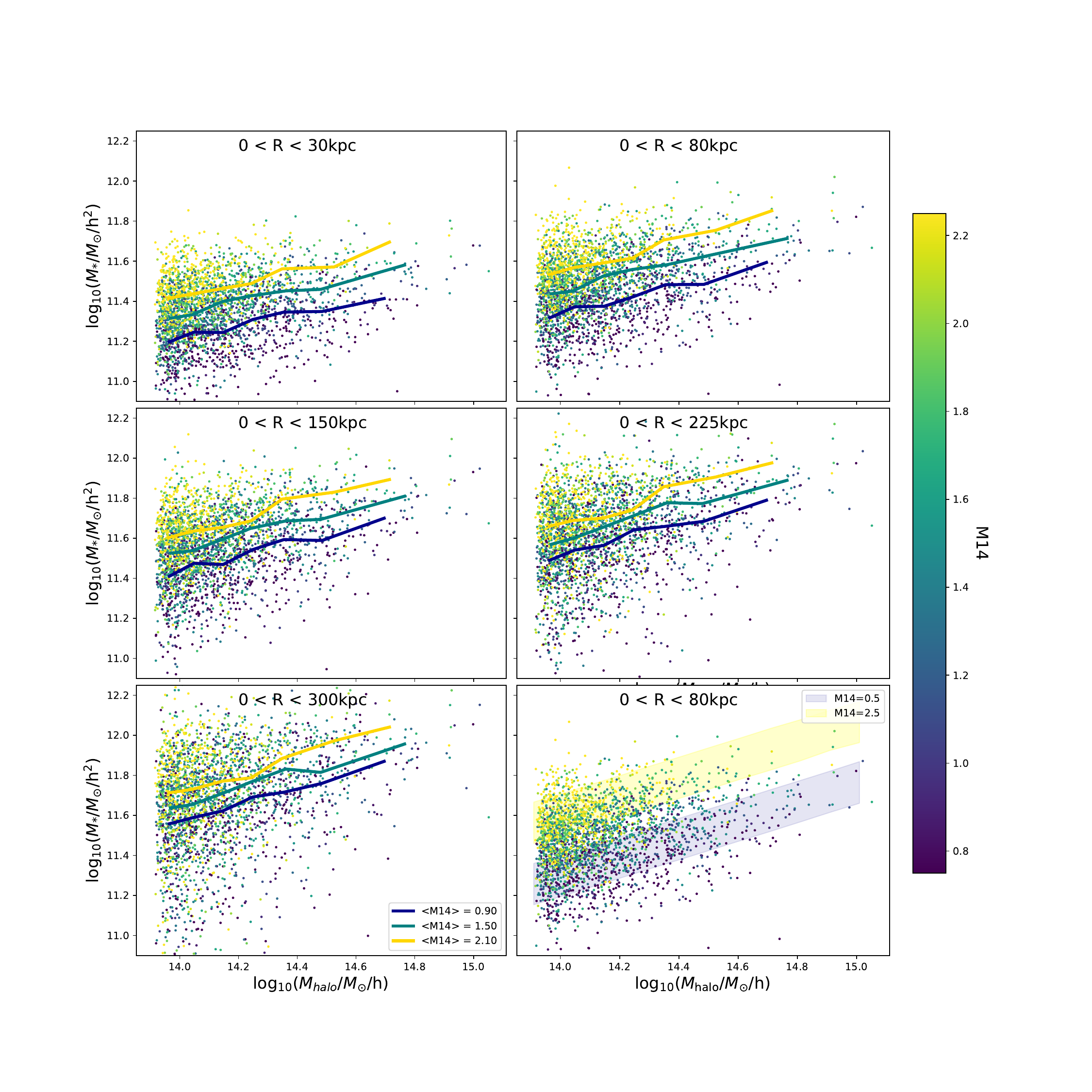}
    \caption{The data used to measure the SMHM relation including the BCG's core. In each sub-figure, we plot the BCG stellar mass against the halo mass. The outer radius used to measure the stellar mass is measured to progressively larger radii from 30\,kpc to 300\,kpc. The solid lines are representative of the median values of three M14 bins, each containing one-third of the data. In the last sub-figure, we overlay the SMHM relation posterior distribution for a value of M14 = 0.5 and M14 = 2.5. Moreover, in each sub-figure the colour is the M14 value of the data shown. As we include light from larger radii, we see that the data becomes more scattered, though the M14-stellar mass stratification continues to persist.} 
    \label{fig:SMHM_core}
\end{figure*}

First, we address the entire BCG+ICL profile and how the SMHM relation varies as we increase the outer radius. Specifically, we use the 0-30\,kpc, 0-80\,kpc, 0-150\,kpc, 0-225\,kpc, and 0-300\,kpc apertures as shown in Figure~\ref{fig:SMHM_core}. We note that the ICL is defined as incorporating all light beyond the transition region ( $r >$ 80\,kpc). However, as we were unsure of whether M14 continues to correlate with BCG+ICL stellar mass out to the full extent of the ICL, we divide the ICL into multiple apertures. 

Figure~\ref{fig:SMHM_core} visualizes the stellar mass within the selected aperture plotted against the halo mass for each annulus, colour coded by M14. In each subplot, we overplot three lines, which are representative of the median of one-third of the data sorted by M14. The median M14 values are 0.90 (dark blue), 1.50 (green) and 2.10 (yellow). Based on Equation~\ref{eq:SMHM_M14}, not accounting for noise, at fixed halo mass, the separation of these lines is given by $\gamma \times 0.6$, where 0.6 is the difference in the average M14 values. Although not shown, we note that the 1$\sigma$ values for these median lines increases with aperture size from 0.1 to 0.2. In the final subplot, we overlay the fits made to Equation~\ref{eq:SMHM_M14} for our observed data as described by the posterior distribution from our MCMC analysis. For example, we fix M14 to the upper and lower M14 values to illustrate our fitting. 

\begin{table*}
\caption{The median and 1$\sigma$ values for each of the measured SMHM relation parameters (given in Equations~\ref{eq:SMHM_M14} and \ref{eq:SMHM}) in each of the different measurement apertures and when M14 is and is not included.}
\centering
\begin{tabular}{ccccccc}
\hline
Data & Inner Radius & Outer Radius & $\alpha$ & $\beta$ & $\gamma$ & $\sigma_{\rm int}$ \\
\hline
Core & 0 & 30 &$0.170 \pm 0.008$ & $0.249 \pm 0.018$ & -- & $0.136 \pm 0.002$ \\
Core & 0 & 80 &$0.325 \pm 0.009$ & $0.321 \pm 0.020$ & -- & $0.147 \pm 0.002$  \\
Core & 0 & 150 &$0.422 \pm 0.009$ & $0.357 \pm 0.020$ & -- & $0.155 \pm 0.002$  \\
Core & 0 & 225 &$0.499 \pm 0.011$ & $0.391 \pm 0.024$ & -- & $0.180 \pm 0.003$  \\
Core & 0 & 300 &$0.561 \pm 0.012$ & $0.405 \pm 0.027$ & -- & $0.208 \pm 0.003$  \\\hline
Core & 0 & 30 &$-0.001 \pm 0.008$ & $0.398 \pm 0.016$ & $0.152 \pm 0.004$ & $0.092 \pm 0.002$  \\
Core & 0 & 80 &$0.153 \pm 0.008$ & $0.464 \pm 0.017$ & $0.153 \pm 0.004$ & $0.103 \pm 0.002$  \\
Core & 0 & 150 &$0.274 \pm 0.010$ & $0.488 \pm 0.019$ & $0.134 \pm 0.004$ & $0.123 \pm 0.002$  \\
Core & 0 & 225 &$0.375 \pm 0.011$ & $0.500 \pm 0.021$ & $0.112 \pm 0.005$ & $0.162 \pm 0.003$  \\
Core & 0 & 300 &$0.456 \pm 0.014$ & $0.495 \pm 0.027$ & $0.094 \pm 0.007$ & $0.197 \pm 0.003$  \\\hline
No Core & 0 & 30 &$0.177 \pm 0.009$ & $0.262 \pm 0.020$ & -- & $0.136 \pm 0.003$ \\
No Core & 30 & 80 &$-0.207 \pm 0.013$ & $0.431 \pm 0.031$ & --& $0.211 \pm 0.004$  \\ 
No Core & 80 & 150 &$-0.262 \pm 0.014$ & $0.351 \pm 0.032$ & -- & $0.221 \pm 0.004$  \\ 
No Core & 150 & 225 &$-0.250 \pm 0.016$ & $0.281 \pm 0.037$ & -- & $0.262 \pm 0.004$  \\ 
No Core & 225 & 300 &$-0.274 \pm 0.020$ & $0.252 \pm 0.047$ & -- &$0.334 \pm 0.006$  \\ \hline
No Core & 0 & 30 &$0.006 \pm 0.008$ & $0.408 \pm 0.017$ & $0.154 \pm 0.004$ & $0.087 \pm 0.003$  \\
No Core & 30 & 80 &$-0.364 \pm 0.015$ & $0.565 \pm 0.029$ & $0.142 \pm 0.007$ & $0.187 \pm 0.004$  \\ 
No Core & 80 & 150 &$-0.314 \pm 0.017$ & $0.389 \pm 0.033$ & $0.044 \pm 0.009$ & $0.219 \pm 0.004$  \\ 
No Core & 150 & 225 &$-0.273 \pm 0.020$ & $0.306 \pm 0.038$ &$0.023 \pm 0.010$ &$ 0.262 \pm 0.004$  \\ 
No Core & 225 & 300 &$-0.281 \pm 0.023$ & $0.229\pm 0.045$ &$0.021 \pm 0.011$ &$0.300 \pm 0.005$  \\ \hline

\end{tabular}
\label{tab:SMHM_Posteriors}
\end{table*}

To quantify the impact of including the ICL and M14 on the parameters of the SMHM relation, we present the median and 1$\sigma$ values for each of the SMHM relation parameters measured from the posterior distribution (both for when M14 is and is not incorporated) in Table~\ref{tab:SMHM_Posteriors}. Additionally, we present an example 2D posterior distribution in Figure~\ref{fig:example} in the Appendix. For consistency, we see that the results in the radial range of $0 < R < 30$\,kpc are in excellent agreement with the results presented in \citet{gol19} and \citet{gol22}. Moreover, these SMHM relation parameters are within 1$\sigma$ of the values of $\beta$ and $\sigma_{\rm int}$ measured in \citet{gol23} (due to the different median subtraction values, we can't compare $\alpha$), which shows our results are not biased by the use of richness-inferred halo masses as opposed to SZ-estimated halo masses. 

\begin{figure}
    \centering
    \includegraphics[width=\columnwidth]{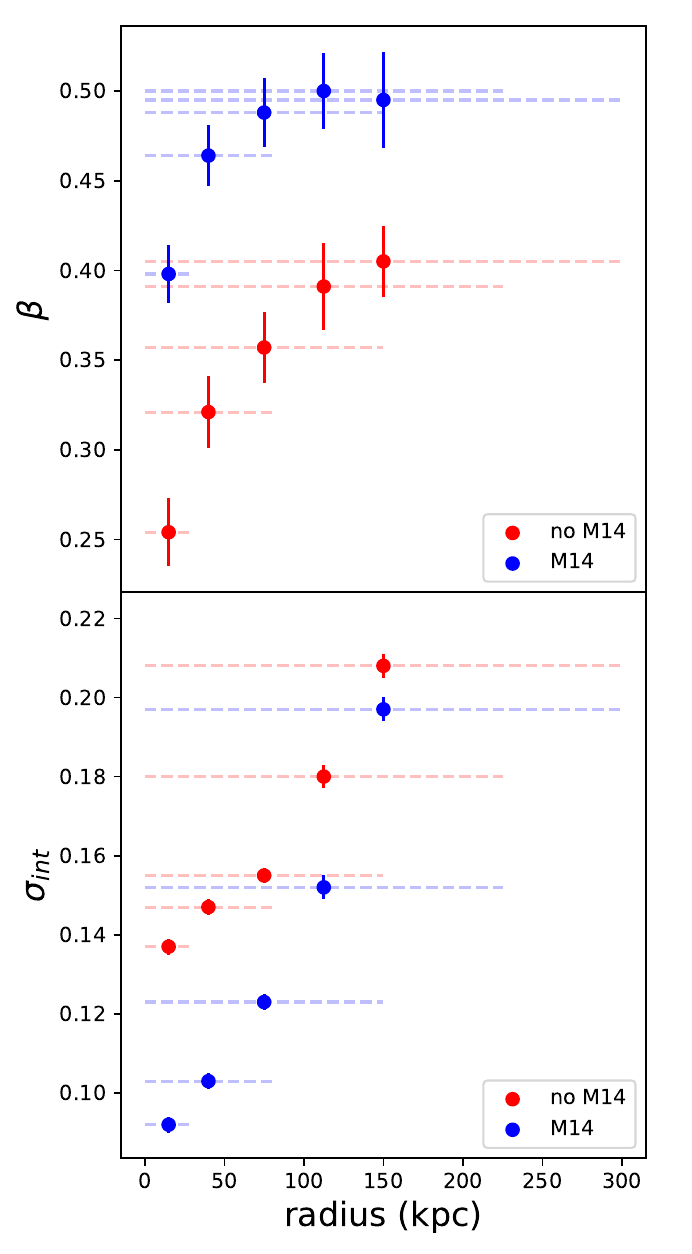}
    \caption{We compare the inferred median and 1$\sigma$ values for $\beta$ and $\sigma_{\rm int}$ measured from the posterior distribution of our MCMC analysis when M14 is (blue) and is not (red) incorporated. The dashed horizontal lines represent the radial range over which the measurement is made, while the solid vertical lines represent the 1$\sigma$ uncertainty from the posterior. M14 is a latent parameter out to large radii in the SMHM relation because $\beta$ increases and $\sigma_{\rm int}$ decreases.}
    \label{fig:SMHM_param_M14}
\end{figure}

In terms of including M14, Figure~\ref{fig:SMHM_param_M14} directly compares how $\beta$ and $\sigma_{\rm int}$ vary with outer radius when M14 is (blue) and is not (red) incorporated. 
Across each aperture, we see clear trends such that the incorporation of M14 increases $\beta$ by $\approx$0.10-0.15 matching prior results \citep{gol18,gol19,gol22}. Moreover, $\beta$ reaches an asymptote at a radius of 150\,kpc, in strong agreement with the trends found in \citet{gol19} and \citet{gol23}. Additionally, the inclusion of the ICL increases $\beta$ by 0.1-0.15 as found in \citet{gol23} (a similar increase to what is found by including M14). As in \citet{gol19}, the value $\beta$ asymptotes at is significantly higher when M14 is accounted for, by approximately 0.1. 

The change of $\sigma_{\rm int}$ is more nuanced. When M14 is incorporated we find that $\sigma_{\rm int}$ decreases, in agreement with the trends for the BCG \citep{gol18,gol19,gol22}. Interestingly, the decrease in $\sigma_{\rm int}$ that results from the inclusion of M14, is reduced from 0.04 dex to 0.02 dex as we extend our measurement to larger radii, which may reflect that M14 is more strongly correlated with the light in the inner portion of the total BCG+ICL profile, as investigated in Section~\ref{subsec:outskirts_core}. 

Additionally, although Figure~\ref{fig:SMHM_param_M14} does not present the $\gamma$ measurements, the median and 1$\sigma$ values are given in Table~\ref{tab:SMHM_Posteriors} and range from 38$\sigma$ to 13$\sigma$ from 0. Thus, the M14-stellar mass correlation persists out to 300\,kpc. However, $\gamma$ decreases as we progress to larger radii, which is shown by the slightly decreasing separation between the solid lines shown in Figure~\ref{fig:SMHM_core}. 

The presence of a non-zero $\gamma$ along with the reduction in $\sigma_{\rm int}$ and increase in $\beta$ allow us to conclude that M14 is a latent parameter when the SMHM relation is measured out to the ICL. For further verification, we randomly reassign the M14 values. If a non-zero value is measured for $\gamma$, this would suggest an intrinsic, non-physical, correlation between these parameters. However, as shown in Table~\ref{tab:SMHM_Posteriors_shuffled}, for each bin, the shuffled $\gamma$ value is equivalent to 0. Moreover, the values of $\alpha$, $\beta$, and $\sigma_{\rm int}$ are in excellent agreement between the versions with the randomly assigned M14 (Table~\ref{tab:SMHM_Posteriors_shuffled}) and when M14 is not incorporated (Table~\ref{tab:SMHM_Posteriors}). Thus, any change in parameter measurements results solely from a correlation between the BCG+ICL stellar mass and M14. Going forward, we include M14 in all measurements discussed in this analysis (though we provide the posteriors when M14 is not accounted for in Tables~\ref{tab:SMHM_Posteriors}, ~\ref{tab:SMHM_Posteriors_PCenlow}, and ~\ref{tab:SMHM_Posteriors_PCenhigh}).

\subsection{Comparison of the SMHM relation of the BCG to the SMHM relation of outskirts}
\label{subsec:outskirts_core}

Although the correlation between the stellar mass and the halo mass increases as the BCG+ICL light profile is extended outward, as discussed in \citet{gol23}, \citet{hua21}, and \citet{kwi24}, the strongest correlation between stellar mass and halo mass has been measured when the core of the BCG is excluded. Here, we investigate the inclusion of M14 and use 5 non-overlapping radial apertures, 0-30\,kpc, 30\,kpc-80\,kpc (representative of the ICL transition region \citep{zha24}), 80\,kpc-150\,kpc, 150\,kpc-225\,kpc, 225\,kpc-300\,kpc. We note that the data set used in this part of the analysis is a slightly different sample as what was used in Section~\ref{subsec:BCG_ICL_SMHM} because we require that the aperture have a non-zero stellar mass in each of the 5 apertures, so there are some clusters where we can measure a non-zero total light profile (starting from the core), but not an individual aperture mass (i.e., there are some clusters where the data may not be deep enough to allow for measurements of the large outer apertures). 

\begin{figure*}
    \centering
    \includegraphics[width=20cm,trim={3cm 3.5cm 1cm 3.5cm},clip]{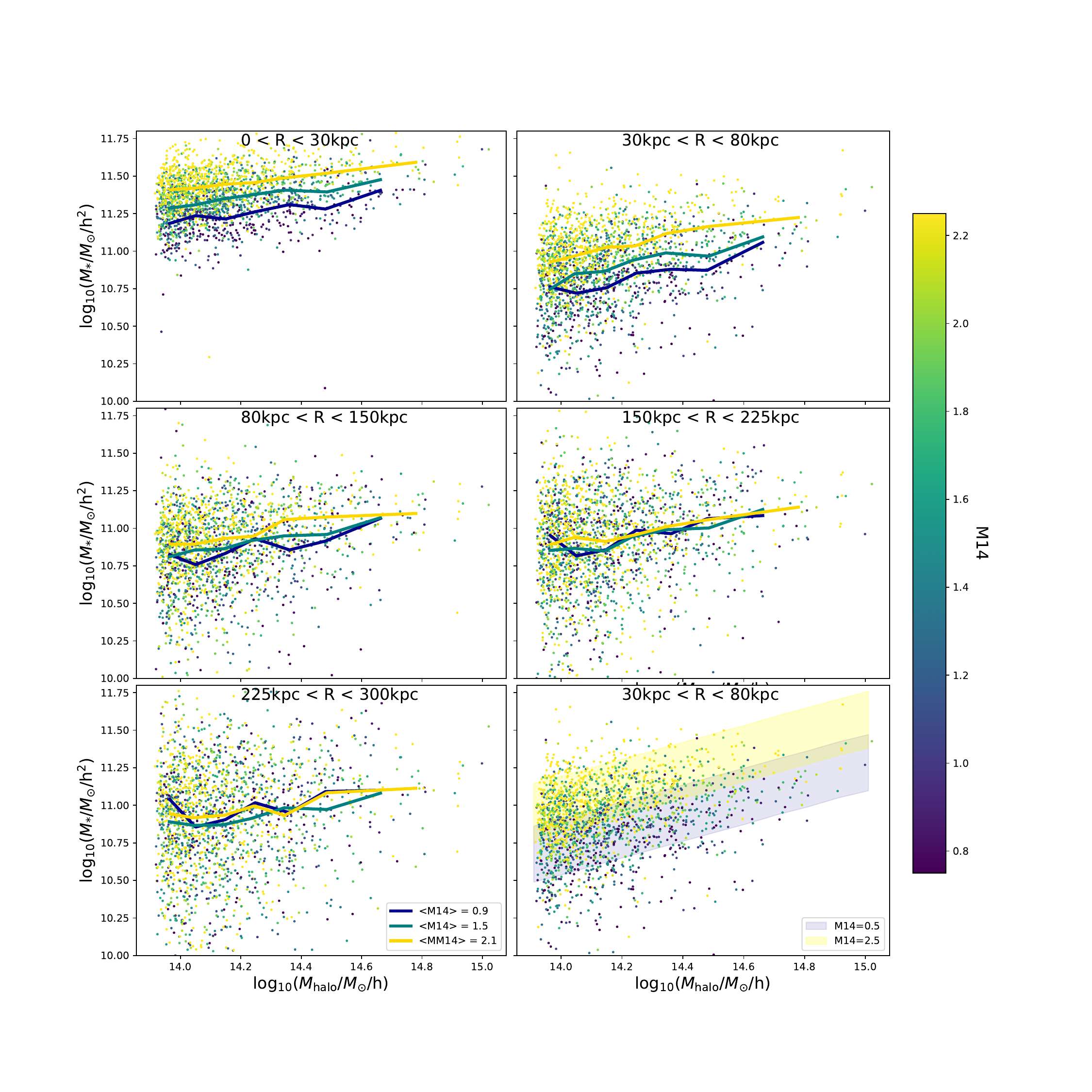}
    \caption{The data used to measure the SMHM relation within unique non-overlapping apertures. In each sub-figure, we plot the stellar mass against the halo mass. The inner and outer radius used to measure the stellar mass progressively increases from 0-30\,kpc to 225\,kpc - 300\,kpc. As we look at light from the ICL, we no longer see a correlation between M14 and stellar mass at fixed halo mass. The solid lines represent the median values of three M14 bins containing one-third of the data each. In the last subfigure, we overlay the results of the posterior distribution for a value of M14 = 0.5 and M14=2.5 in the 30\,kpc < R < 80\,kpc aperture.}
    \label{fig:SMHM_nocore}
\end{figure*}

In Figure~\ref{fig:SMHM_nocore}, we plot the data used to measure the SMHM relation, the stellar mass vs. the halo mass coloured using M14. Although the stellar mass range contained within each aperture varies, we fix the y-axis, to highlight the changing value. In each subplot, we overlay the median values of stellar mass and halo mass for the three previously defined M14 bins. Although not shown, the 1$\sigma$ value increases from 0.1 to 0.3 dex as the aperture moves outward. In the final subplot, we again overlay the fits made to Equation~\ref{eq:SMHM_M14} for our observed data, which is given by the posterior distribution from our MCMC analysis, where we fix M14 to the upper and lower values to illustrate our fitting. The stellar mass-M14 correlation persists strongly in both the 0-30\,kpc and 30\,kpc-80\,kpc apertures. Beyond this aperture, both the overall distribution and any trend with M14 becomes dominated by the noise, as demonstrated by the solid lines which begin to overlap. 

\begin{figure}
    \centering
    \includegraphics[width=7.5cm]{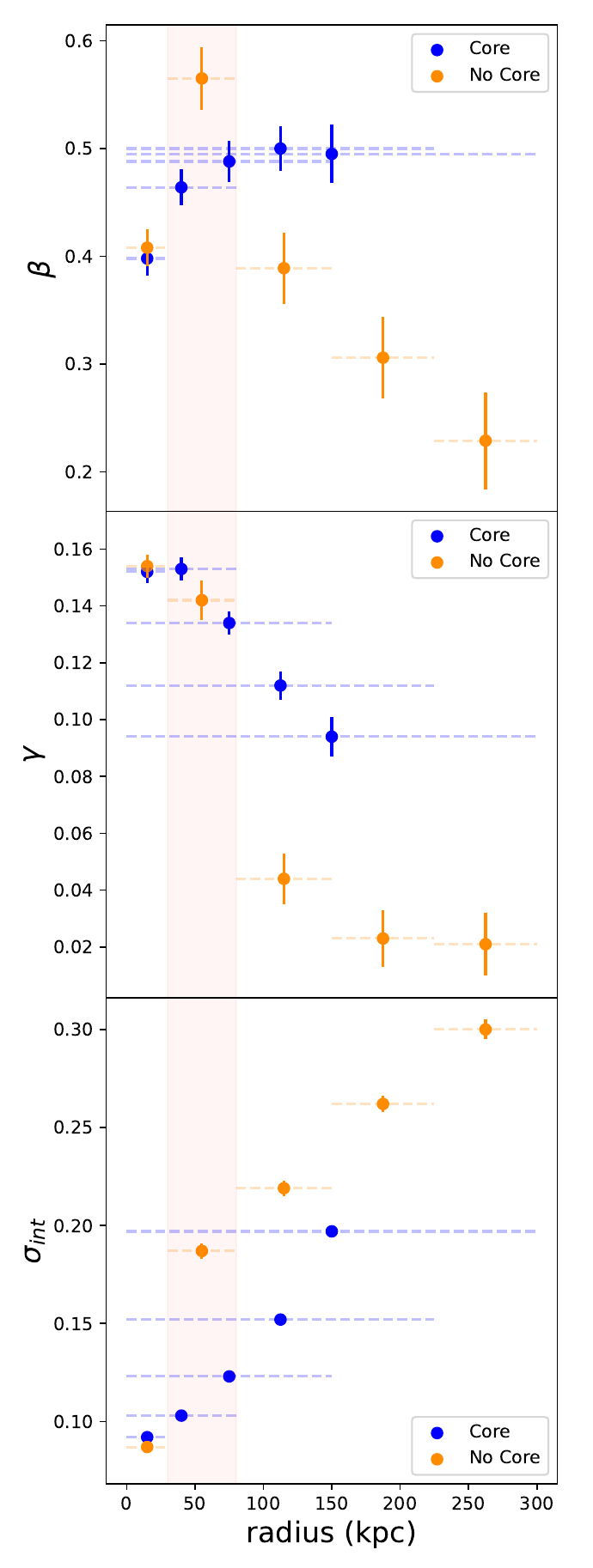}
    \caption{We compare the inferred median values derived from the posterior distribution for $\beta$, $\gamma$, and $\sigma_{\rm int}$ when the core of the BCG is (blue) and is not (orange) incorporated. The dashed horizontal lines show the radial range where each measurement is made, while the solid vertical lines are the 1$\sigma$ values based on the posterior distribution. We see that the strongest correlations (in terms of $\beta$, $\gamma$, and $\sigma_{\rm int}$) occur in the radial range 30-80\,kpc, which is shaded in light red, and that the correlation with M14 becomes weaker when looking at only light beyond 150\,kpc.}
    \label{fig:SMHM_param_core}
\end{figure} 
We visually compare how using the core of the BCG impacts the $\beta$, $\gamma$, and $\sigma_{\rm int}$ parameters in Figure~\ref{fig:SMHM_param_core}. For $\beta$, we measure the steepest slope in the radial range 30-80\,kpc. This slope that is almost 0.07 (15\%) steeper than any version measured when the light from the core of the BCG is used, which supports that the light contained within the ICL is strongly correlated with the cluster's halo mass. Moreover, this is similar to the result found in \citet{gol23} that the slope is peaked in the radial aperture of 30\,kpc-50\,kpc and also agrees with results presented in \citet{hua21} that find the strongest correlation with halo mass using the radial range of 50\,kpc - 100\,kpc. Additionally, we note that the light from the outskirts (beyond 150\,kpc) is almost as strongly correlated with the dark matter halo mass as the BCG's core, though with a significantly larger scatter.   

The trend measured for $\sigma_{\rm int}$ is nearly identical to what is found in \citet{gol23}, where $\sigma_{\rm int}$ is significantly lower when the light from the core of the BCG is included. Since BCG cores can be treated as a standard candle \citep[e.g.,][]{lau07}, this is unsurprising. However, we remind the reader that our estimate for $\sigma_{\rm int}$ is an upper limit because we assume the same uncertainty on the stellar mass for each of the radial bins, calibrated for the BCG 0-30\,kpc region. If we are underestimating the uncertainty in the stellar mass of the ICL, then we would be overestimating $\sigma_{\rm int}$. 

Lastly, the most interesting and surprising result comes from $\gamma$, which is statistically more than 5$\sigma$ from 0 for the 0-30\,kpc, 30\,kpc-80\,kpc, and 80\,kpc-150\,kpc apertures (the maximum value is 38$\sigma$ and then decreases to 5$\sigma$ as we move to farther out apertures). This means that the stellar mass-M14 correlation persists out to 150\,kpc, though its strength decreases drastically beyond 80\,kpc. Moreover, the value of $\gamma$ is statistically equivalent in the BCG's core and the BCG+ICL transition region (30\,kpc-80\,kpc). Since M14 is independent of the light from this radial regime, this may offer further support that the light in the transition region is also growing as a result of BCG hierarchical assembly. Additionally, we note that at large radii (r > 150\,kpc), we detect no correlation between stellar mass and M14 (the significance of $\gamma$ is less than 2$\sigma$), which may lead to the decrease in $\gamma$ measured when the BCG's core is included and shows that for that dataset, the correlation between M14 and stellar mass is coming from the inner region (between 0-80\,kpc or 0-150\,kpc) of the BCG+ICL system.

\section{Physical Meaning of the Correlation between the Magnitude Gap and the ICL}
\label{sec:M14_ICL}

\subsection{The SMHM Relation}
Figures~\ref{fig:SMHM_core} and ~\ref{fig:SMHM_nocore} illustrate the connection between M14 and the ICL as inferred from the SMHM relation. Table~\ref{tab:SMHM_Posteriors} and Figure~\ref{fig:SMHM_param_core} quantify that the ICL and M14 correlate, such that for certain radial bins, the inclusion of M14 yields a tighter SMHM relation. However, as shown in Figure~\ref{fig:SMHM_param_core}, the strength of this correlation weakens as we extend the ICL out to larger apertures and stops being statistically significant at radii beyond 150\,kpc. 

Figure~\ref{fig:SMHM_param_core} and Table~\ref{tab:SMHM_Posteriors} also highlight that the strength of the correlation between the BCG+ICL light/stellar mass and M14 is within 1$\sigma$ for the 0-30\,kpc, 0-80\,kpc, and 30\,kpc - 80\,kpc apertures. Moreover, both when the core is and is not included, $\gamma$ declines at radii beyond 80\,kpc. While the $\gamma$ parameters are similar, the radial range that features the strongest correlation between the stellar mass and the underlying halo mass is the BCG+ICL transition region (30\,kpc - 80\,kpc). Although this aperture does not have the lowest $\sigma_{\rm int}$, the value ($\sigma_{\rm int}$ = 0.187) is in agreement with the values assumed when M14 is not incorporated ($\sim 0.15$). Therefore, from a statistical standpoint, there is a trade-off between the lower $\sigma_{\rm int}$ when the core is included and the higher $\beta$ when it is not. Additionally, we note that the inclusion of M14 decreases $\sigma_{\rm int}$ by 0.034 dex and increases $\beta$ by 0.13, in the transition region, which is the largest change in both parameters for one of the unique apertures when incorporating M14 into the SMHM relation. 

We remind the reader that the stellar mass measured between 30\,kpc-80\,kpc and M14 are independent. Thus, this strong $\gamma$ parameter informs the BCG+ICL system's formation. Following the results of \citet{gol18}, M14 can be thought of as a tracer for hierarchical growth, such that clusters with larger M14 values are more evolved and possibly older systems. Thus, the presence of such a strong M14 stratification in the BCG+ICL transition region suggests that the hierarchical growth of the BCG is not confined to the core, but rather extends to the BCG+ICL transition regime. Moreover, it is plausible that a greater fraction of the stellar mass growth from this hierarchical merging is deposited within the transition region, which may lead to the strong correlation. Assuming this scenario, the BCG+ICL transition region then grows predominately through the major/minor mergers of the BCG, while the ICL may grow through tidal stripping of satellite galaxies (as shown by the detected colour gradient \citep[e.g.,][]{mon18,dem18,con19,gol23}). Moreover, because we use radial apertures to separate the ICL and BCG, it is possible that due to projection effects, the correlation with the ICL is even stronger as some fraction of the ICL is contained inside 30\,kpc.  

Although redshift evolution is beyond the scope of this analysis, we present an alternative formation scenario. \citet{gol22} find the stellar mass contained within the core of the BCG is constant over this redshift range. Thus, \citet{gol22} posit that the growth must be occurring at larger radial apertures, possibly within the ICL. However, this growth was not detected in \citet{gol23}, though that analysis focused on the 50\,kpc - 300\,kpc aperture (and did not account for M14), which may have washed out any evolution signal due to the increase in noise. Thus, the correlation with M14 may suggest recent growth is not occurring in the far outskirts of the ICL, but rather in the BCG+ICL transition region, as \citet{zha24} find preliminary evidence for.   

\subsection{The ICL Fraction vs. Magnitude Gap}
\label{subsec:ICLfrac}
Along with the correlation between the ICL and M14 highlighted via the SMHM relation, we measure the ICL fraction as a function of M14 since both parameters have been posited to correlate with the dynamical state of the cluster. We measure three unique ICL fractions defined by the following equations.
\begin{equation}
\label{eq:ICLfrac1}
    \mathrm{ICL~Fraction}_\mathrm{30 < r < 150}  = \frac{\mathrm{Lum}_\mathrm{ICL} (30<r<150) }{\mathrm{Lum}_\mathrm{ICL+Sat} (30<r<150) } 
\end{equation}
Equation~\ref{eq:ICLfrac1} is the ratio of the ICL light to the total light within the cluster (ICL + satellite galaxies, excluding the BCG) measured between 30\,kpc and 150\,kpc. 
\begin{equation}
\label{eq:ICLfrac2}
    \mathrm{BCG+ICL~Fraction}_\mathrm{0 < r < 150}  = \frac{\mathrm{Lum}_\mathrm{BCG+ICL} (0<r<150) }{\mathrm{Lum}_\mathrm{BCG+ICL+Sat} (0<r<150) } 
\end{equation}
Equation~\ref{eq:ICLfrac2} compares the BCG+ICL light to the total light of the cluster measured within 150\,kpc.  
\begin{equation}
\label{eq:ICLfrac3}
    \mathrm{ICL~Fraction}_\mathrm{0 < r < 150}  = \frac{\mathrm{Lum}_\mathrm{ICL} (30<r<150) }{\mathrm{Lum}_\mathrm{BCG+ICL+Sat} (0<r<150) } 
\end{equation}
Equation~\ref{eq:ICLfrac3} compares the ICL light to the total light in the cluster within 150\,kpc. 

For this portion of the analysis, we stack the measurements of the ICL in 5 M14 bins (20\% quintiles), and then measure the mean ICL fraction and M14 in each bin. For this analysis, we do not further bin the data by halo mass as \citet{zha24}, using the same ICL measurements, detect no correlation between ICL fraction and richness. This lack of correlation between ICL fraction and halo mass is also found in observations from \citet{rag23} and simulations presented in \citet{con14}, \citet{bro24}, and \citet{con24}. Moreover, we remind the reader that Figure~\ref{fig:M14_Mhalo} shows that there is no correlation between M14 and halo mass in our measurements since the median value is constant across the entire halo mass range. Additionally, we note that the median halo mass differs by less than 1$\sigma$ within each of the M14 bins.

\begin{figure}
    \centering
    \includegraphics[width=7.5cm]{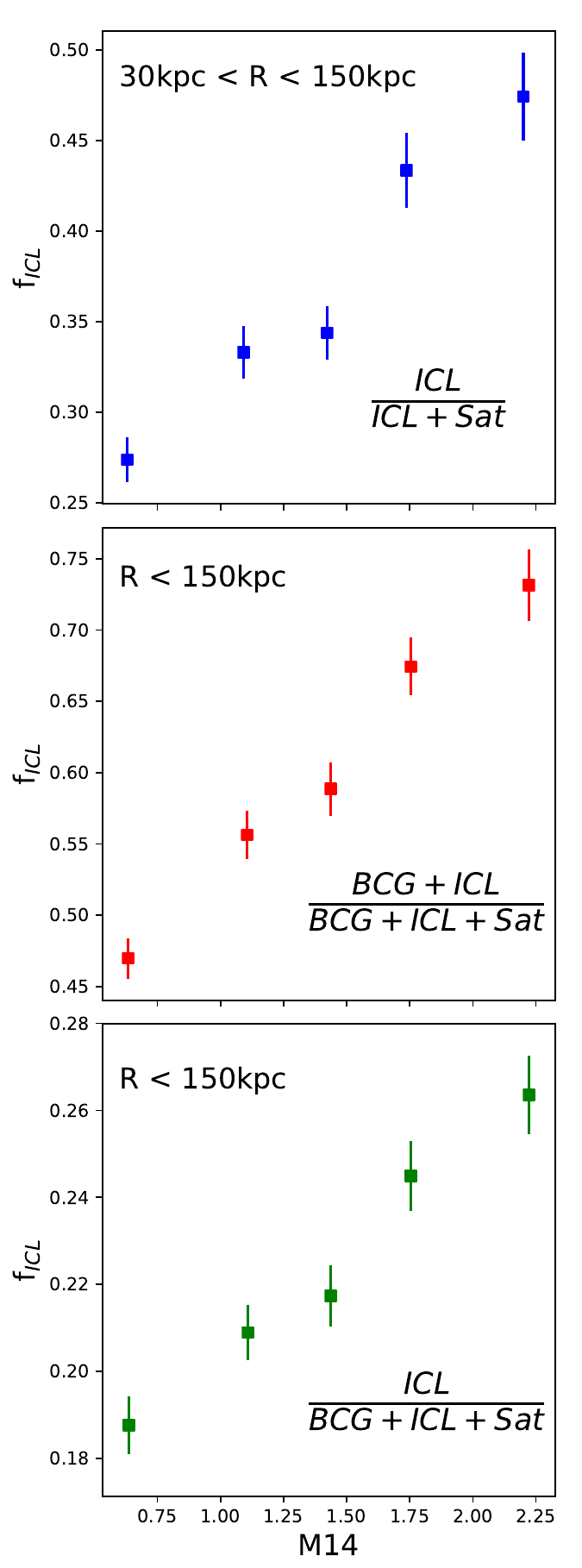}
    \caption{Using stacked measurements of the ICL in 5 M14 bins (20\% quintiles), we compare the ICL fraction to M14. The upper panel is a measure of the ICL fraction in the radial range 30\,kpc < R < 150\,kpc given by Equation~\ref{eq:ICLfrac1}. The middle panel represents the BCG+ICL fraction measured within 150\,kpc given by Equation~\ref{eq:ICLfrac2}. The lower panel represence the ICL fraction measured within 150\,kpc given by Equation~\ref{eq:ICLfrac3}. The errorbars are representative of a boostrap uncertainty from 100,000 draws.}
    \label{fig:ICL_fraction}
\end{figure}

As shown in Figure~\ref{fig:ICL_fraction}, we for the first time using observations find that clusters characterized by a large M14 have significantly larger ICL fractions. Each ICL fraction increases by between 50-75\% when comparing low M14 (M14 < 1.0) and high M14 (M14 > 2.0) clusters. As a result of hierarchical merging, we expect a fraction of stars will be ejected into the ICL while another fraction will increase the BCG stellar mass. Thus, the M14 and ICL fraction correlation directly results from BCG hierarchical assembly. Therefore, it follows that the increase in the fraction as a function of M14 is much larger when the BCG's light is included (comparing the middle and lower panels), as the BCG's stellar mass increases due to hierarchical assembly. Moreover, following the interpretation of M14 from \citet{gol18} and M12 from \citet{vit18} -- clusters with larger M14 values form earlier and have had more time to evolve -- we posit that this may mean that clusters characterized by larger ICL fractions are dynamically older and that as clusters evolve, the BCG+ICL system (and as a result M14 and ICL fraction) grows. Additionally, though not shown, similar to \citet{zha24}, we find that the ICL fraction is dependent on the choice of aperture and decreases as the ICL is extended out to 300\,kpc

Although early works \citep{rud06, con14, Cui14} showed evidence of a correlation between the ICL fraction and the cluster's dynamical state, this has become a topic of focus in many recent works in both simulations and small observational samples. Using semi-analytic models, \citet{con23} show that the ICL fraction positively correlates with halo concentration, a tracer of cluster dynamical state, such that more relaxed clusters have larger ICL fractions. Similarly, \citet{bro24}, using the Magneticum Pathfinder simulation, find that the ICL fraction negatively correlates with the mass fraction of the eighth subhalo, another tracer of cluster dynamical state, such that more relaxed clusters have larger BCG+ICL fractions.  Furthermore, using The 300 Project, \citet{con24} find that the ICL fraction negatively correlates with both the offset of the centre of mass from the BCG and the subhalo mass fraction, both dynamical state tracers, so that more relaxed clusters form earlier and have a larger ICL fraction.

While a consensus exists in simulations and semi-analytic models, there are discrepancies in observations.  Using six Hubble Frontier Field clusters \citet{mon18} find that their most relaxed cluster has a larger ICL fraction (measured using elliptical apertures between 50\,kpc and 120\,kpc) than the least relaxed cluster. Similarly, \citet{rag23} find a modest positive correlation between the ICL fraction and the early-type galaxy fraction, an observational tracer of more dynamically evolved and relaxed clusters. However, in contrast, \citet{jim18}, \citet{dup22}, and \citet{jim24} have found the ICL fraction has an inverse correlation with the dynamical state of the cluster, such that relaxed clusters are characterized by low ICL fractions, while merging clusters are characterized by larger ICL fractions. 

As previously noted, the magnitude gap is also viewed as a tracer for cluster dynamical state. In the context of observations of fossil group galaxies, \citet{zar15} suggest that systems characterized by larger M12 values are dynamically relaxed. Additionally, recent observational work by \citet{cas24} find that applying a high M14 criteria (M14 > 1.25) better identifies dynamically relaxed clusters. This trend is further supported by cosmological hydrodynamic simulations from \citet{yoo24}, which find that clusters with larger M12 values are more relaxed. Moreover, \citet{yoo24} also find a strong correlation between M12 and BCG+ICL fraction, similar to what is shown in the middle panel of Figure~\ref{fig:ICL_fraction}. Although we use M14, as shown in \citet{gol18}, there is no statistical difference between these parameters due to the measurement uncertainty. Thus, the results shown in Figure~\ref{fig:ICL_fraction} are in excellent agreement with the simulations from \citet{yoo24}. 

This literature comparison is not perfect. For example, \citet{bro24} look at lower mass clusters while \citet{zar15}, \citet{rag23}, and \citet{yoo24} focus on lower halo mass groups, \citet{mon18} have only 6 clusters, \citet{jim18}, \citet{dup22}, and \citet{jim24} use X-ray measurements, measure the ICL using CICLE \citep[e.g.,][]{jim18}, and do not estimate the ICL fraction within a fixed aperture, and \citet{cas24} also analyze an X-ray sample and do not consider the ICL. However, the excellent agreement with the simulations from \citet{yoo24} and agreement with the previously described works strongly support that both ICL fraction and M14 correlate with dynamical state such that larger M14 values and higher ICL fractions are typical of dynamically relaxed systems. However, given this contradiction to results from \citet{jim18}, \citet{dup22}, and \citet{jim24} we posit an explanation. \citet{zha24} find that the ICL fraction is dependent on the choice of radial aperture, and thus it is possible that the measurements from \citet{jim18}, \citet{dup22}, and \citet{jim24} are not directly comparable to our own. However, further investigation, using simulations and consistent ICL measurement methods and definitions are required to gain greater insight into this apparent discrepancy.  

\section{Conclusion}
\label{sec:conclusion}
In this work, our analysis focuses on characterizing how the ICL and M14 are related using the SMHM relation as both measurements are related to BCG hierarchical assembly. We use DES Year 6 data to characterize the SMHM relation over the redshift range $0.2 < z < 0.6$. Throughout the analysis we treat the BCG and ICL as a linked system using radial cuts to define certain regions. We summarize the primary results of this analysis as follows. 
\begin{itemize}
    \item When the BCG+ICL stellar mass is measured from the core out to progressively larger apertures, $\gamma$, the correlation between stellar mass and M14 at fixed halo mass is non-zero out to large radii. Using our Bayesian infrastructure, we find that the inclusion of M14 increases $\beta$ by 0.1-0.15 and decreases $\sigma_{\rm int}$ by 0.01-0.04 dex (depending on the choice of aperture). Thus, M14 is a latent variable in the SMHM relation. However, the impact of M14 decreases at larger radii. 
    \item When measuring the SMHM relation in unique, non-overlapping apertures, we find that the strongest correlation between stellar mass and halo mass is in the 30\,kpc - 80\,kpc bin. This result follows from the ``two-phase'' formation of BCGs and agrees with results from \citet{hua21} and \citet{gol23}. Moreover, $\beta$ increases by 30\% with the inclusion of M14. 
    \item When using unique apertures, the stellar mass - M14 correlation is statistically significant in the 0-30\,kpc, 30\,kpc - 80\,kpc, and 80\,kpc - 150\,kpc apertures. $\gamma$ decreases significantly beyond 80\,kpc. Most importantly, $\gamma$ is equivalent in the 0-30\,kpc, 0-80\,kpc, and 30\,kpc - 80\,kpc apertures. Since M14 contains no information about the light in the 30\,kpc - 80\,kpc aperture, this correlation highlights that the stellar mass in this region likely grows as a result of BCG hierarchical assembly. Thus, it is plausible that the BCG+ICL transition region grows as a result of mergers, while the outer ICL grows through processes such as tidal stripping. 
    \item We for the first time observationally measure a strong correlation between the mean ICL fraction and M14 value such that clusters characterized by a larger M14 (M14 > 2.0) have ICL fractions that are more than 50\% higher than clusters with a small M14 (M14 < 1.0). As M14 grows as a result of mergers, this correlation follows from BCG hierarchical assembly.  
\end{itemize}
Going forward, there remain many research questions we aim to explore to improve our observational understanding and characterisation of the BCG+ICL. In particular, understanding how the BCG+ICL stellar mass and radial distribution evolves with redshift and varies with wavelength. In future analysis using the \textit{Euclid} Mission (as demonstrated by \citet{klu24}) and the Rubin Observatory's LSST (as discussed in \citet{mon19review} and \citet{bro24}), we plan to extend this type of analysis out to higher redshifts ($z\approx 1.5)$) and into the IR regime to investigate the formation and evolution of the BCG+ICL system and how its correlation with halo mass evolves. Moreover, given the correlation between M14 and ICL fraction, we suggest that characterizing large M14 systems in \textit{Euclid} and LSST, or targetting them with JWST, may provide information about the earliest appearances of the ICL. Additionally, we aim to determine whether the correlation between M14 and BCG stellar mass persists out to high redshift and whether this is intrinsic (i.e., is it in place prior to ex-situ growth) or results entirely from hierarchical formation. 

\section*{Acknowledgements}

\textit{Author Contributions}: We would like to acknowledge everyone who made this work possible. JBGM developed the Bayesian infrastructure that was used in this analysis, made the measurements of each parameter, did the analysis of the ICL and M14, and was the principle author responsible for writing the manuscript. YZ developed the infrastructure used to measure the ICL in the DES data, produced the ICL light profiles used in this analysis, provided guidance, and aided in paper writing and discussion. RLCO provided useful insights throughout this work and aided in paper writing and discussion. BY assisted with our use of the DES data and construction of the ICL pipeline. MESP and MH provided detailed and useful feedback as DES internal reviewers. The remaining authors have made contributions to this paper that include, but are not limited to, the construction of DECam and other aspects of collecting the data; data processing and calibration; developing broadly used methods, codes, and simulations; running the pipelines and validation tests; and promoting the science analysis. 

JBGM would like to thank the anonymous referee for all of their help and feedback, which significantly strengthened this manuscript. JBGM would like to thank Nina Hatch, the members of the NottICL research group, and the participants in the ISSI ICL working group for many useful discussions about the nature of ICL and measurement techniques. JBGM gratefully acknowledges support from the Leverhulme Trust.  

\textit{Funding}: Funding for the DES Projects has been provided by the U.S. Department of Energy, the U.S. National Science Foundation, the Ministry of Science and Education of Spain, the Science and Technology Facilities Council of the United Kingdom, the Higher Education Funding Council for England, the National Center for Supercomputing Applications at the University of Illinois at Urbana-Champaign, the Kavli Institute of Cosmological Physics at the University of Chicago, the Center for Cosmology and Astro-Particle Physics at the Ohio State University,the Mitchell Institute for Fundamental Physics and Astronomy at Texas A\&M University, Financiadora de Estudos e Projetos, Funda{\c c}{\~a}o Carlos Chagas Filho de Amparo {\`a} Pesquisa do Estado do Rio de Janeiro, Conselho Nacional de Desenvolvimento Cient{\'i}fico e Tecnol{\'o}gico and the Minist{\'e}rio da Ci{\^e}ncia, Tecnologia e Inova{\c c}{\~a}o, the Deutsche Forschungsgemeinschaft and the Collaborating Institutions in the Dark Energy Survey. 

The Collaborating Institutions are Argonne National Laboratory, the University of California at Santa Cruz, the University of Cambridge, Centro de Investigaciones Energ{\'e}ticas, Medioambientales y Tecnol{\'o}gicas-Madrid, the University of Chicago, University College London, the DES-Brazil Consortium, the University of Edinburgh, the Eidgen{\"o}ssische Technische Hochschule (ETH) Z{\"u}rich, Fermi National Accelerator Laboratory, the University of Illinois at Urbana-Champaign, the Institut de Ci{\`e}ncies de l'Espai (IEEC/CSIC), the Institut de F{\'i}sica d'Altes Energies, Lawrence Berkeley National Laboratory, the Ludwig-Maximilians Universit{\"a}t M{\"u}nchen and the associated Excellence Cluster Universe, the University of Michigan, NSF NOIRLab, the University of Nottingham, The Ohio State University, the University of Pennsylvania, the University of Portsmouth, SLAC National Accelerator Laboratory, Stanford University, the University of Sussex, Texas A\&M University, and the OzDES Membership Consortium.

Based in part on observations at NSF Cerro Tololo Inter-American Observatory at NSF NOIRLab (NOIRLab Prop. ID 2012B-0001; PI: J. Frieman), which is managed by the Association of Universities for Research in Astronomy (AURA) under a cooperative agreement with the National Science Foundation.

The DES data management system is supported by the National Science Foundation under Grant Numbers AST-1138766 and AST-1536171. The DES participants from Spanish institutions are partially supported by MICINN under grants PID2021-123012, PID2021-128989 PID2022-141079, SEV-2016-0588, CEX2020-001058-M and CEX2020-001007-S, some of which include ERDF funds from the European Union. IFAE is partially funded by the CERCA program of the Generalitat de Catalunya.

We acknowledge support from the Brazilian Instituto Nacional de Ci\^enciae Tecnologia (INCT) do e-Universo (CNPq grant 465376/2014-2).

This document was prepared by the DES Collaboration using the resources of the Fermi National Accelerator Laboratory (Fermilab), a U.S. Department of Energy, Office of Science, Office of High Energy Physics HEP User Facility. Fermilab is managed by Fermi Forward Discovery Group, LLC, acting under Contract No. 89243024CSC000002.

\section*{Data Availability}

The data underlying this article will be shared on reasonable request to the corresponding author.

\begin{appendix}
\section{Verification Tests}

In Tables~\ref{tab:SMHM_Posteriors_PCenlow} and ~\ref{tab:SMHM_Posteriors_PCenhigh}, we present the posterior distributions when we divide our sample into two subsets based on the redMaPPer central galaxy probability ($P_{\rm Cen}$). We use two samples with $P_{\rm Cen} < 0.60$ (Table~\ref{tab:SMHM_Posteriors_PCenlow}) and $P_{\rm Cen} > 0.99$ (Table~\ref{tab:SMHM_Posteriors_PCenhigh}). When comparing these posterior distributions to those presented in Table~\ref{tab:SMHM_Posteriors}, we see that all measurements are consistent within 1$\sigma$. This provides additional evidence that our results are not biased by miscentring or incorrect BCG identification, as BCGs with low $P_{\rm Cen}$ are more likely to be miscentred or misidentified as the BCG.  
\begin{table*}
\caption{Posterior Distribution Results PCen < 60\%}
\centering
\begin{tabular}{ccccccc}
\hline
Data & Inner Radius & Outer Radius & $\alpha$ & $\beta$ & $\gamma$ & $\sigma_{\rm int}$ \\
\hline
Core & 0 & 30 &$0.178 \pm 0.024$ & $0.260 \pm 0.053$ & -- & $0.136 \pm 0.006$ \\
Core & 0 & 80 &$0.323 \pm 0.026$ & $0.308 \pm 0.057$ & -- & $0.143 \pm 0.007$ \\
Core & 0 & 150 &$0.434 \pm 0.027$ & $0.360 \pm 0.059$ & -- & $0.150 \pm 0.007$ \\
Core & 0 & 225 &$0.527 \pm 0.031$ & $0.425 \pm 0.069$ & -- & $0.177 \pm 0.008$ \\
Core & 0 & 300 &$0.596 \pm 0.036$ & $0.449 \pm 0.079$ & -- & $0.204 \pm 0.009$ \\\hline
Core & 0 & 30 &$0.023 \pm 0.023$ & $0.426 \pm 0.046$ & $0.146 \pm 0.011$ & $0.095 \pm 0.006$ \\
Core & 0 & 80 &$0.181 \pm 0.026$ & $0.460 \pm 0.052$ & $0.133 \pm 0.013$ & $0.111 \pm 0.007$ \\
Core & 0 & 150 &$0.312 \pm 0.029$ & $0.489 \pm 0.056$ & $0.112 \pm 0.014$ & $0.129 \pm 0.007$ \\
Core & 0 & 225 &$0.421 \pm 0.035$ & $0.511 \pm 0.069$ & $0.091 \pm 0.017$ & $0.166 \pm 0.008$ \\
Core & 0 & 300 &$0.533 \pm 0.043$ & $0.511 \pm 0.083$ & $0.057 \pm 0.021$ & $0.201 \pm 0.009$ \\\hline
No Core & 0 & 30 &$0.180 \pm 0.027$ & $0.264 \pm 0.061$ & -- & $0.137 \pm 0.007$ \\
No Core & 30 & 80 &$-0.208 \pm 0.037$ & $0.406 \pm 0.084$ & --& $0.206 \pm 0.010$ \\ 
No Core & 80 & 150 &$-0.224 \pm 0.043$ & $0.419 \pm 0.097$ & -- & $0.231 \pm 0.011$ \\ 
No Core & 150 & 225 &$-0.175 \pm 0.049$ & $0.455 \pm 0.111$ & -- & $0.273 \pm 0.013$ \\ 
No Core & 225 & 300 &$-0.190 \pm 0.063$ & $0.440 \pm 0.143$ & -- &$0.352 \pm 0.016$ \\ \hline
No Core & 0 & 30 &$0.021 \pm 0.027$ & $0.412 \pm 0.052$ & $0.146 \pm 0.013$ & $0.096 \pm 0.007$ \\
No Core & 30 & 80 &$-0.319 \pm 0.045$ & $0.506 \pm 0.087$ & $0.101 \pm 0.022$ & $0.195 \pm 0.010$ \\ 
No Core & 80 & 150 &$-0.261 \pm 0.052$ & $0.435 \pm 0.099$ & $0.029 \pm 0.026$ & $0.230 \pm 0.011$ \\ 
No Core & 150 & 225 &$-0.179 \pm 0.060$ & $0.465 \pm 0.116$ &$0.005 \pm 0.030$ &$ 0.272 \pm 0.013$ \\ 
No Core & 225 & 300 &$-0.204 \pm 0.069$ & $0.336 \pm 0.132$ &$0.005 \pm 0.035$ &$0.316 \pm 0.015$ \\ \hline

\end{tabular}
\label{tab:SMHM_Posteriors_PCenlow}
\end{table*}

\begin{table*}
\caption{Posterior Distribution Results PCen > 99\%}
\centering
\begin{tabular}{ccccccc}
\hline
Data & Inner Radius & Outer Radius & $\alpha$ & $\beta$ & $\gamma$ & $\sigma_{\rm int}$ \\
\hline
Core & 0 & 30 &$0.183 \pm 0.014$ & $0.285 \pm 0.032$ & -- & $0.133 \pm 0.004$ \\
Core & 0 & 80 &$0.333 \pm 0.015$ & $0.349 \pm 0.035$ & -- & $0.146 \pm 0.004$ \\
Core & 0 & 150 &$0.424 \pm 0.016$ & $0.375 \pm 0.036$ & -- & $0.155 \pm 0.005$ \\
Core & 0 & 225 &$0.503 \pm 0.018$ & $0.412 \pm 0.041$ & -- & $0.176 \pm 0.005$ \\
Core & 0 & 300 &$0.577 \pm 0.020$ & $0.450 \pm 0.046$ & -- & $0.199 \pm 0.006$ \\\hline
Core & 0 & 30 &$-0.015 \pm 0.014$ & $0.396 \pm 0.027$ & $0.157 \pm 0.007$ & $0.091 \pm 0.004$ \\
Core & 0 & 80 &$0.141 \pm 0.016$ & $0.472 \pm 0.029$ & $0.159 \pm 0.008$ & $0.099 \pm 0.005$ \\
Core & 0 & 150 &$0.259 \pm 0.017$ & $0.487 \pm 0.032$ & $0.139 \pm 0.009$ & $0.120 \pm 0.005$ \\
Core & 0 & 225 &$0.362 \pm 0.021$ & $0.504 \pm 0.039$ & $0.117 \pm 0.010$ & $0.156 \pm 0.005$ \\
Core & 0 & 300 &$0.462 \pm 0.024$ & $0.526 \pm 0.045$ & $0.094 \pm 0.012$ & $0.188 \pm 0.006$ \\\hline
No Core & 0 & 30 &$0.189 \pm 0.015$ & $0.283 \pm 0.036$ & -- & $0.134 \pm 0.005$ \\
No Core & 30 & 80 &$-0.201 \pm 0.023$ & $0.463 \pm 0.055$ & --& $0.213 \pm 0.007$ \\ 
No Core & 80 & 150 &$-0.259 \pm 0.023$ & $0.398 \pm 0.055$ & -- & $0.214 \pm 0.007$ \\ 
No Core & 150 & 225 &$-0.257 \pm 0.027$ & $0.319 \pm 0.066$ & -- & $0.256 \pm 0.008$ \\ 
No Core & 225 & 300 &$-0.276 \pm 0.034$ & $0.295 \pm 0.083$ & -- &$0.329 \pm 0.010$ \\ \hline
No Core & 0 & 30 &$0.005 \pm 0.016$ & $0.387 \pm 0.029$ & $0.149 \pm 0.008$ & $0.090 \pm 0.005$ \\
No Core & 30 & 80 &$-0.388 \pm 0.027$ & $0.578 \pm 0.051$ & $0.153 \pm 0.014$ & $0.186 \pm 0.007$ \\ 
No Core & 80 & 150 &$-0.316 \pm 0.030$ & $0.425 \pm 0.056$ & $0.045 \pm 0.016$ & $0.212 \pm 0.007$ \\ 
No Core & 150 & 225 &$-0.259 \pm 0.035$ & $0.318 \pm 0.066$ &$0.001 \pm 0.018$ &$ 0.255 \pm 0.008$ \\ 
No Core & 225 & 300 &$-0.247 \pm 0.039$ & $0.270 \pm 0.075$ &$0.001 \pm 0.021$ &$0.285 \pm 0.009$ \\ \hline

\end{tabular}
\label{tab:SMHM_Posteriors_PCenhigh}
\end{table*}

In Table~\ref{tab:SMHM_Posteriors_shuffled}, we present the posterior distributions when M14 is randomly assigned, removing any existing correlation between M14 and aperture stellar mass. For each version, we find that $\gamma$ is equivalent to 0 and that the measured posteriors are in excellent agreement with the values provided in Table~\ref{tab:SMHM_Posteriors} when M14 is not accounted for. The discrepancy between the shuffled results (Table~\ref{tab:SMHM_Posteriors_shuffled}) and those of the real data (Table~\ref{tab:SMHM_Posteriors}) along with the nested nature of our MCMC model further highlight that M14 is indeed a latent parameter, the correlation between aperture stellar mass and M14 is real, and including M14 yields a better fit to the underlying observational data.   

\begin{table*}
\caption{Posterior Distribution Results Shuffled}
\centering
\begin{tabular}{ccccccc}
\hline
Data & Inner Radius & Outer Radius & $\alpha$ & $\beta$ & $\gamma$ & $\sigma_{\rm int}$ \\
\hline
Core & 0 & 30 &$0.167 \pm 0.011$ & $0.257 \pm 0.018$ & $0.002 \pm 0.004$  & $0.139 \pm 0.002$ \\
Core & 0 & 80 &$0.321 \pm 0.012$ & $0.318 \pm 0.020$ & $0.002 \pm 0.005$ & $0.147 \pm 0.002$  \\
Core & 0 & 150 &$0.427 \pm 0.012$ & $0.358 \pm 0.021$ & $-0.002 \pm 0.005$ & $0.155 \pm 0.002$  \\
Core & 0 & 225 &$0.503 \pm 0.014$ & $0.393 \pm 0.024$ & $-0.002 \pm 0.006$ & $0.180 \pm 0.003$  \\
Core & 0 & 300 &$0.561 \pm 0.017$ & $0.398 \pm 0.028$ & $-0.001 \pm 0.007$ & $0.210 \pm 0.003$  \\\hline
No Core & 0 & 30 &$0.175 \pm 0.013$ & $0.263 \pm 0.021$ & $0.001 \pm 0.005$ & $0.136 \pm 0.003$ \\
No Core & 30 & 80 &$-0.210 \pm 0.019$ & $0.424 \pm 0.031$ & $0.000 \pm 0.008$& $0.212 \pm 0.004$  \\ 
No Core & 80 & 150 &$-0.255 \pm 0.019$ & $0.351 \pm 0.032$ & $-0.004 \pm 0.008$ & $0.221 \pm 0.004$  \\ 
No Core & 150 & 225 &$-0.238 \pm 0.023$ & $0.275 \pm 0.038$ & $-0.009 \pm 0.009$ & $0.263 \pm 0.005$  \\ 
No Core & 225 & 300 &$-0.249 \pm 0.027$ & $0.208 \pm 0.043$ & $-0.005 \pm 0.011$ &$0.300 \pm 0.005$  \\ \hline

\end{tabular}
\label{tab:SMHM_Posteriors_shuffled}
\end{table*}

\section{Example Posterior}
\begin{figure}
    \centering
    \includegraphics[width=7.5cm]{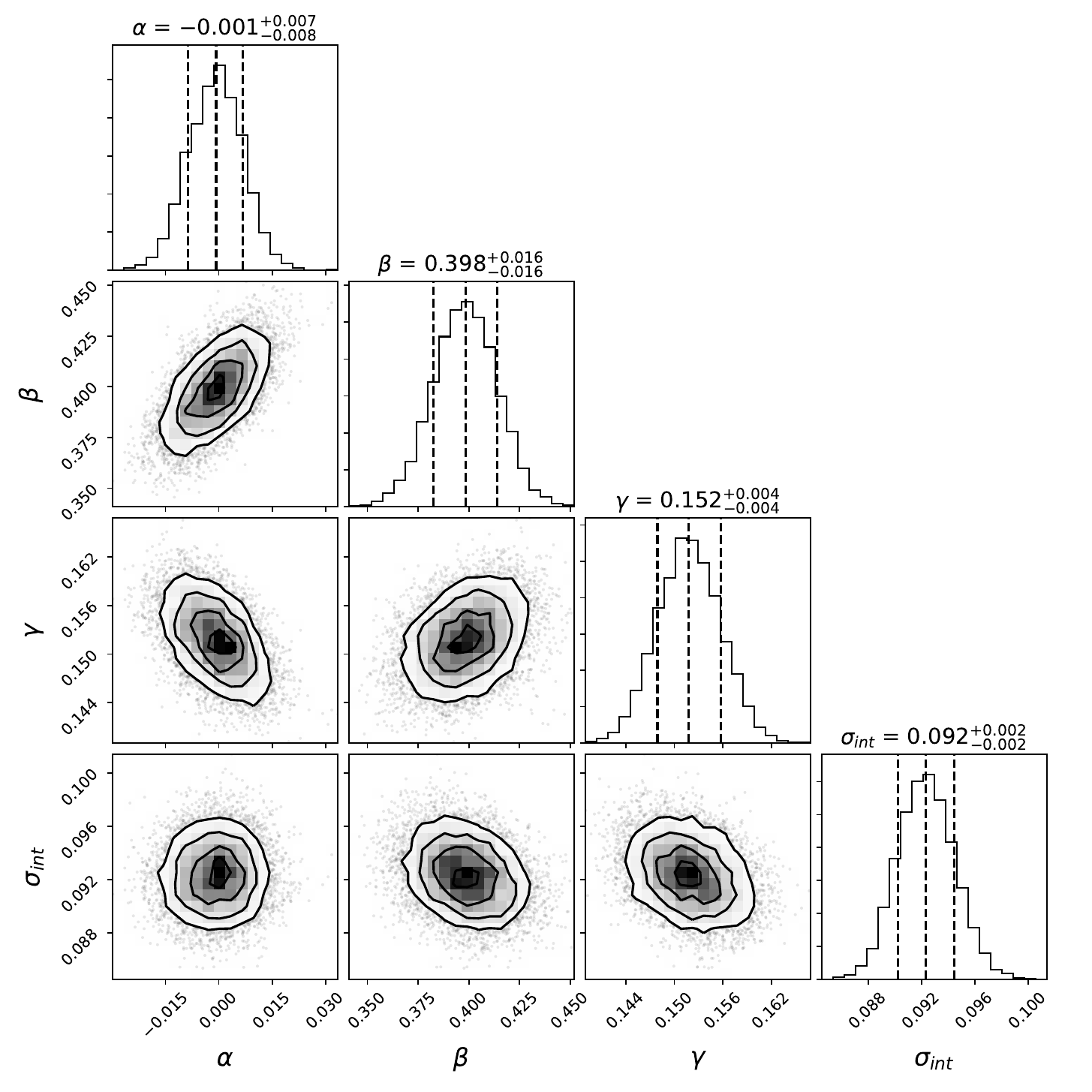}
    \caption{An example 2D posterior distribution for the BCG region, including M14. }
    \label{fig:example}
\end{figure}

In the MCMC analysis described in Section~\ref{sec:model}, we note that our output is given by the 2D posterior distribution. Since we run a number of MCMC analyses in this work, we choose not to display each individual posterior distribution because they are quite similar and can be more concisely summarized in Tables~\ref{tab:SMHM_Posteriors}, \ref{tab:SMHM_Posteriors_PCenlow}, \ref{tab:SMHM_Posteriors_PCenhigh}, \ref{tab:SMHM_Posteriors_shuffled}. However, here, we present an example posterior distribution for the BCG+ICL region with M14 included. Figure~\ref{fig:example} highlights the covariance that exists between $\alpha$ and $\beta$, which is reduced by removing the median values, and also that the values presented in each table represent the median and 1$\sigma$ values from the posterior distribution.

\end{appendix}

\section*{Affiliations}
{\small
$^{1}$ School of Physics and Astronomy, University of Nottingham, Nottingham, NG7 2RD, UK\\
$^{2}$ NSF's National Optical-Infrared Astronomy Research Laboratory, 950 Cherry Ave., Tucson, Az 85719, USA\\
$^{3}$ Observat\'orio Nacional, Rua Gal. Jos\'e Cristino 77, Rio de Janeiro, RJ - 20921-400, Brazil\\
$^{4}$ Fermi National Accelerator Laboratory, P. O. Box 500, Batavia, IL 60510, USA\\
$^{5}$ Hamburger Sternwarte, Universit\"{a}t Hamburg, Gojenbergsweg 112, 21029 Hamburg, Germany\\
$^{6}$ Wits Centre for Astrophysics, School of Physics, University of Witwatersrand, Private Bag 3, 2050, Johannesburg, South Africa\\
$^{7}$ Astrophysics Research Centre, School of Mathematics, Statistics, and Computer Science, University of KwaZulu-Natal, Westville Campus, Durban 4041, South Africa\\
$^{8}$ Laborat\'orio Interinstitucional de e-Astronomia - LIneA, Rua Gal. Jos\'e Cristino 77, Rio de Janeiro, RJ - 20921-400, Brazil\\
$^{9}$ Department of Physics, University of Michigan, Ann Arbor, MI 48109, USA\\
$^{10}$ Institute of Cosmology and Gravitation, University of Portsmouth, Portsmouth, PO1 3FX, UK\\
$^{11}$ Department of Physics \& Astronomy, University College London, Gower Street, London, WC1E 6BT, UK\\
$^{12}$ Instituto de Astrofisica de Canarias, E-38205 La Laguna, Tenerife, Spain\\
$^{13}$ Institut de F\'{\i}sica d'Altes Energies (IFAE), The Barcelona Institute of Science and Technology, Campus UAB, 08193 Bellaterra (Barcelona) Spain\\
$^{14}$ Centre for Extragalactic Astronomy, Durham University, South Road, Durham, DH1 3LE, UK\\
$^{15}$ Centro de Investigaciones Energ\'eticas, Medioambientales y Tecnol\'ogicas (CIEMAT), Madrid, Spain\\
$^{16}$ Department of Physics, IIT Hyderabad, Kandi, Telangana 502285, India\\
$^{17}$ Jet Propulsion Laboratory, California Institute of Technology, 4800 Oak Grove Dr., Pasadena, CA 91109, USA\\
$^{18}$ Institute of Theoretical Astrophysics, University of Oslo. P.O. Box 1029 Blindern, NO-0315 Oslo, Norway\\
$^{19}$ Kavli Institute for Cosmological Physics, University of Chicago, Chicago, IL 60637, USA\\
$^{20}$ Instituto de Fisica Teorica UAM/CSIC, Universidad Autonoma de Madrid, 28049 Madrid, Spain\\
$^{21}$ Department of Physics and Astronomy, University of Pennsylvania, Philadelphia, PA 19104, USA\\
$^{22}$ University Observatory, Faculty of Physics, Ludwig-Maximilians-Universit\"at, Scheinerstr. 1, 81679 Munich, Germany\\
$^{23}$ Center for Astrophysical Surveys, National Center for Supercomputing Applications, 1205 West Clark St., Urbana, IL 61801, USA\\
$^{24}$ Department of Astronomy, University of Illinois at Urbana-Champaign, 1002 W. Green Street, Urbana, IL 61801, USA\\
$^{25}$ School of Mathematics and Physics, University of Queensland, Brisbane, QLD 4072, Australia\\
$^{26}$ Santa Cruz Institute for Particle Physics, Santa Cruz, CA 95064, USA\\
$^{27}$ Center for Cosmology and Astro-Particle Physics, The Ohio State University, Columbus, OH 43210, USA\\
$^{28}$ Department of Physics, The Ohio State University, Columbus, OH 43210, USA\\
$^{29}$ Center for Astrophysics $\vert$ Harvard \& Smithsonian, 60 Garden Street, Cambridge, MA 02138, USA\\
$^{30}$ Australian Astronomical Optics, Macquarie University, North Ryde, NSW 2113, Australia\\
$^{31}$ Lowell Observatory, 1400 Mars Hill Rd, Flagstaff, AZ 86001, USA\\
$^{32}$ LPSC Grenoble - 53, Avenue des Martyrs 38026 Grenoble, France\\
$^{33}$ Instituci\'o Catalana de Recerca i Estudis Avan\c{c}ats, E-08010 Barcelona, Spain\\
$^{34}$ Department of Physics, Carnegie Mellon University, Pittsburgh, Pennsylvania 15312, USA\\
$^{35}$ Kavli Institute for Particle Astrophysics \& Cosmology, P. O. Box 2450, Stanford University, Stanford, CA 94305, USA\\
$^{36}$ SLAC National Accelerator Laboratory, Menlo Park, CA 94025, USA\\
$^{37}$ Department of Physics, Northeastern University, Boston, MA 02115, USA\\
$^{38}$ Physics Department, Lancaster University, Lancaster, LA1 4YB, UK\\
$^{39}$ Computer Science and Mathematics Division, Oak Ridge National Laboratory, Oak Ridge, TN 37831\\
$^{40}$ Argone National Laboratory, 9700 S. Cass Avenue, Lemont, IL 60439, USA\\
$^{41}$ Cerro Tololo Inter-American Observatory, NSF's National Optical-Infrared Astronomy Research Laboratory, Casilla 603, La Serena, Chile\\
$^{42}$ Department of Astronomy, University of California, Berkeley, 501 Campbell Hall, Berkeley, CA 94720, USA\\
$^{43}$ Lawrence Berkeley National Laboratory, 1 Cyclotron Road, Berkeley, CA 94720, USA\\
$^{44}$ School of Physics and Astronomy, University of Southampton, Southampton, SO17 1BJ, UK\\
}

\label{lastpage}
\end{document}